\documentclass [aps,12pt,aps,a4paper,prb]{revtex4}
\input epsf
\usepackage{graphicx}
\usepackage{amssymb}
\usepackage{latexsym}

\begin{document}
\title{Origin of translocation barriers for polyelectrolyte chains}
\author{{\bf Rajeev Kumar\footnote[1]{Current Address: Materials Research Laboratory, University of California, Santa Barbara, CA} and M.Muthukumar \footnote[2]{To whom any correspondence should be addressed, Email : muthu@polysci.umass.edu}}}

\affiliation{\it Dept. of Polymer Science \& Engineering, Materials Research Science \& Engineering Center,\\
 University of Massachusetts, Amherst, MA-01003, USA.}
\date{\today}

\maketitle

\vskip0.1cm
\begin{center}
ABSTRACT
\end{center}
\vskip0.1cm

For single-file translocations of a charged macromolecule through a narrow pore, the crucial step of arrival of an end at the pore suffers from free energy barriers, arising from changes in intrachain electrostatic interaction, distribution of ionic clouds and solvent molecules, and conformational entropy of the chain. All contributing factors to the barrier in the initial stage of translocation are evaluated by using the self-consistent field theory for the polyelectrolyte and the coupled Poisson-Boltzmann description for ions, without radial symmetry. The barrier is found to be essentially entropic, due to conformational changes. For moderate and high salt concentrations, the barriers for the polyelectrolyte chain are quantitatively equivalent to that of uncharged self-avoiding walks. Electrostatic effects are shown to increase the free energy barriers, but only slightly. The degree of ionization, electrostatic interaction strength, decreasing salt concentration and the solvent quality all result in increases in the barrier. 
\clearpage
\section{INTRODUCTION}
\setcounter {equation} {0}
Translocation\cite{cellbook,translocationbook,kasianowicz96,kasianowicz99,branton00,smeets06,butler06,
shklovskii06,muthureview07,muthu07,oukhaled_prl08,oukhaled_epl08,
sung96,muthu99,muthuentropy,muthu04,uiuc06,dimarzio97,dimarzio02,kardar01,andrew07,forrey07} of single polyelectrolyte molecules through narrow pores is one of the most fundamental processes encountered in many biological processes\cite{cellbook,translocationbook} and technological\cite{kasianowicz96,kasianowicz99,branton00,smeets06,butler06,shklovskii06,muthureview07,muthu07,
oukhaled_prl08,oukhaled_epl08} applications. An inevitable need to understand the translocation phenomenon has driven 
the scientific community to study both natural\cite{kasianowicz96,kasianowicz99,branton00,smeets06,butler06,muthureview07,shklovskii06} as well as synthetic\cite{muthureview07,muthu07,oukhaled_prl08,oukhaled_epl08} polyelectrolytes. In these studies, an external driving force 
is used to carry out a successful polyelectrolyte translocation, which is typically due to an applied electric field\cite{kasianowicz96,kasianowicz99,branton00,smeets06,butler06,shklovskii06,muthureview07,muthu07,
oukhaled_prl08,oukhaled_epl08} and in some cases, arise due to an osmotic imbalance (confinement-driven translocation\cite{cellbook,translocationbook,sung96,muthu99,muthuentropy,muthu04,uiuc06}). In addition to the complications coming from an intricate coupling between the 
short range excluded volume interactions and the long-range electrostatics, a satisfactory description of the translocation of a polyelectrolyte chain must take into account many extra factors. In general, translocation of a polyelectrolyte chain from one confined space to the other through a narrow pore may be affected by the dielectric mismatch between the interior and exterior of the confining 
membrane\cite{shklovskii06,oukhaled_prl08,oukhaled_epl08,parsegian69}, nature of the pore\cite{muthureview07,sung96,muthu99,muthuentropy}, 
surface of the confining membrane\cite{muthureview07}, electro-osmotic flow through the transmembrane 
pore\cite{andrew07}, and semi-flexibility of the chain\cite{forrey07}. Furthermore, depending on the area of cross-section of the 
pore, the polyelectrolyte may undergo translocation either as a single-file with linear conformations or as multiply folded conformation. 

Independent of the nature of the driving force and any additional factor mentioned above that may affect translocation, the single-file translocation is envisioned as a two step process\cite{muthureview07}. In the first step, one end of the chain arrives at the pore entrance\cite{muthureview07,shklovskii06,oukhaled_prl08,oukhaled_epl08} and, in the second step, the chain is threaded\cite{muthu07,dimarzio97,dimarzio02,sung96,muthu99,muthuentropy,kardar01} through the pore from one side to the other. It has been recognized  that both steps are associated with entropic barriers, with the first barrier associated with the loss of translational entropy of chain ends and the second barrier with reduction in conformational entropy of the chain.

The extent of the entropic barrier due to conformational changes among various contributing factors  for the experimentally relevant polyelectrolytes is not known. In fact, it is known\cite{rajeev08} that the conformational entropy of a confined polyelectrolyte is only a weak contributor to the free energy where translational entropy of small ions and molecules (counterions, coions, and solvent) is dominant. Reorganization of counterion clouds around deforming polyelectrolyte chains can also contribute significantly to the free energy. Furthermore, the intrachain electrostatic repulsion can stiffen the polymer enabling an easier access to the pore entrance. It is therefore of interest to assess the relative magnitudes of various contributing factors to the free energy barrier associated with the translocation of a flexible polyelectrolyte molecule.

\begin{figure}[ht!]
\vspace*{-0.95cm}
\begin{center}
$\begin{array}{c@{\hspace{.001in}}c@{\hspace{.001in}}}
\includegraphics[width=3in,height=3in]{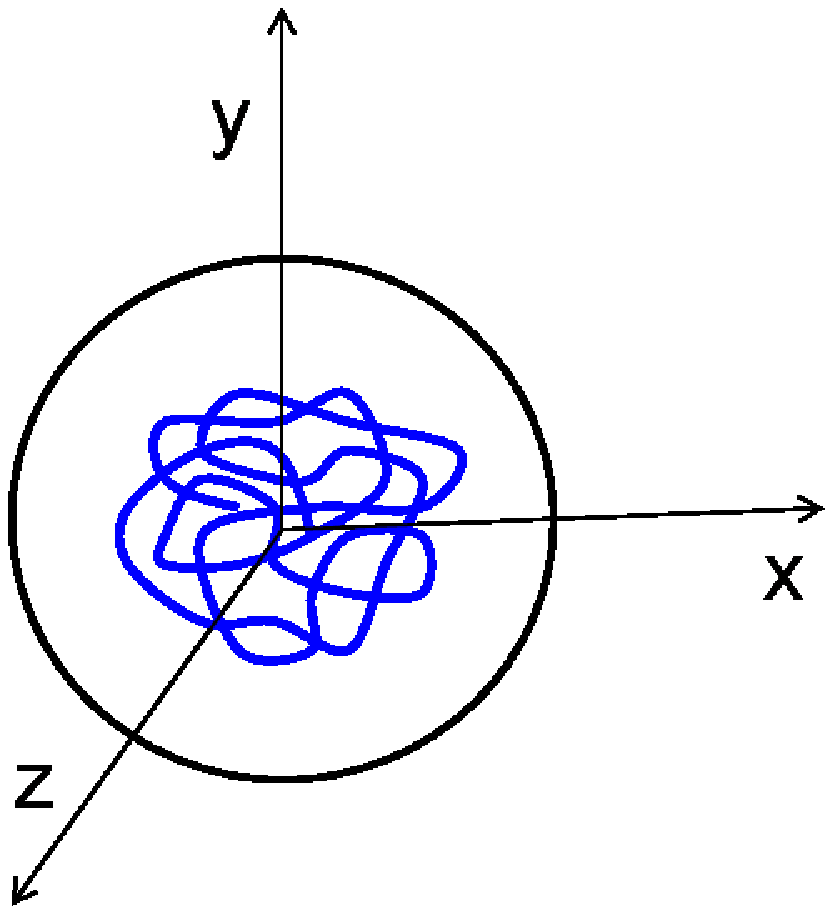}  &
\includegraphics[width=3in,height=3in]{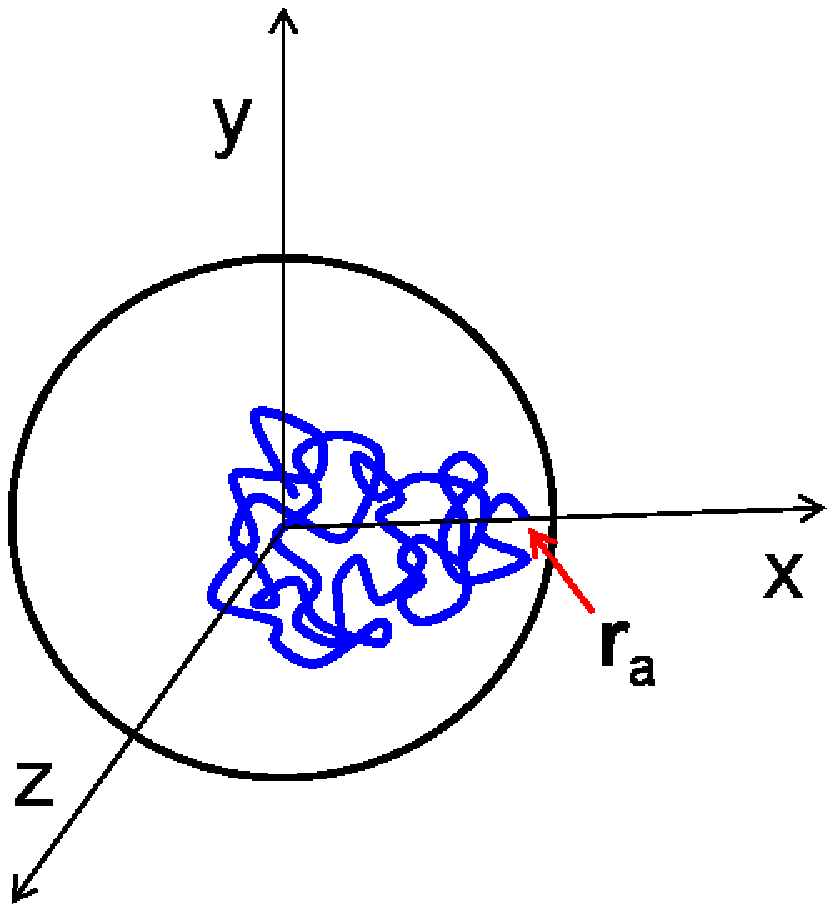} \\
\textbf{(a)} & \textbf{(b)}
\end{array}$
\newline

$\begin{array}{c@{\hspace{.001in}}c@{\hspace{.001in}}}
 \includegraphics[width=3in,height=3in]{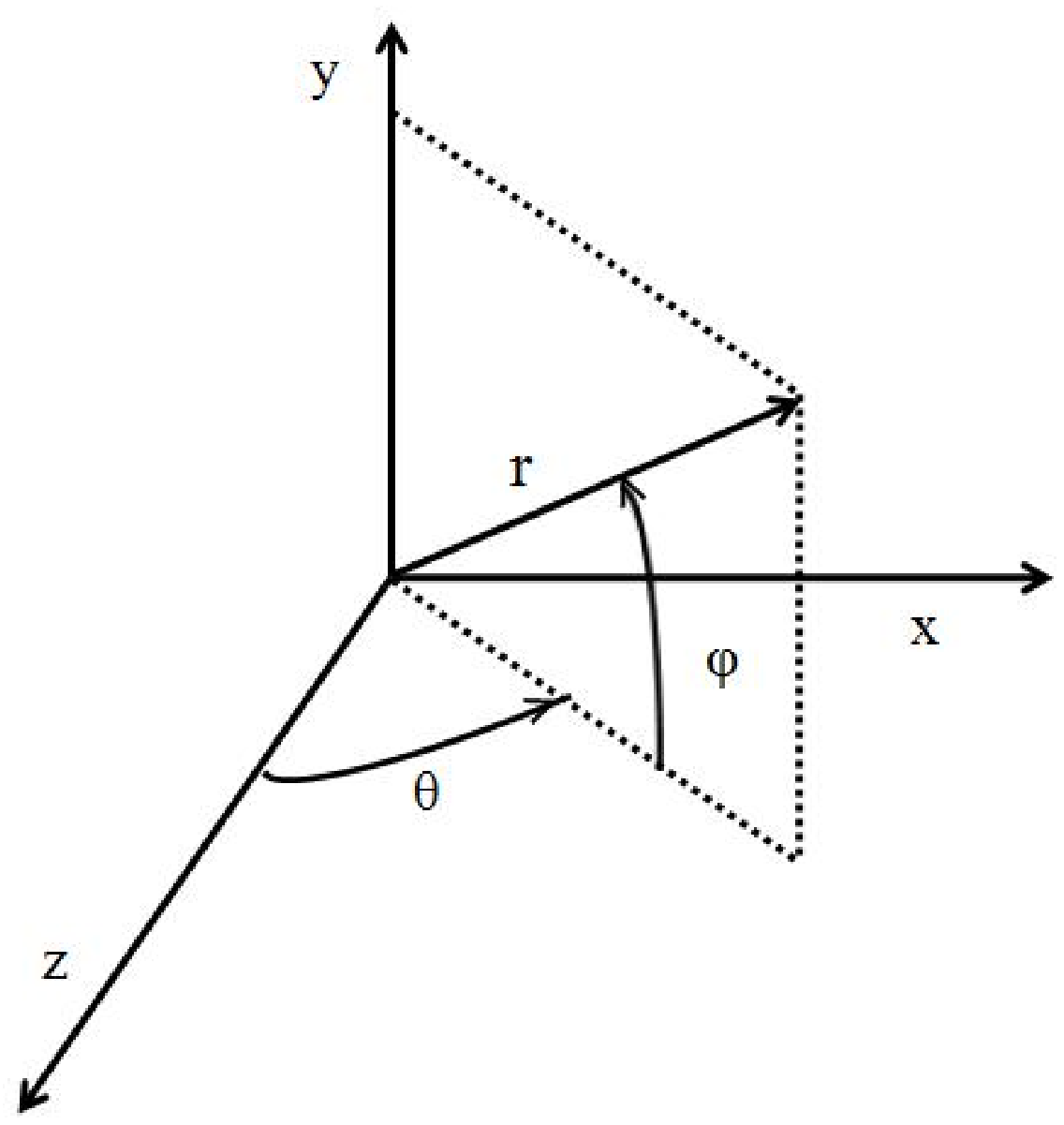} &  \\
 \textbf{(c)} &  
\end{array}$
\vspace*{-1.0cm}
\newline
\end{center}
\caption{Cartoons of the chain in the ``free ends'' (1a) and ``end-fixed'' (1b) states. Spherical coordinates used to solve the SCFT equations are shown in Fig. 1c.} \label{fig:cartoons2}
\end{figure}

Although most of the theoretical works on the entropic barrier for translocation has focused on the threading (second step), the largest 
part of the barrier is actually associated with the first step of localizing one end of the chain at the pore 
entrance\cite{muthureview07}. In this work, we focus on the first step in confinement- driven translocation 
involving single flexible polyelectrolyte chain trying to get out of the confining spherical cavity through a pore on the surface 
as a ``single-file'' and provide a quantitative description of the free energy barriers for 
the chain end to find the pore. The translocation barrier for the first step is estimated by the free energy difference between ``end-fixed'' (one end fixed near the pore) and ``free ends'' (confined chain, which is 
free to move inside the cavity) equilibrium states of the chain (Fig. ~\ref{fig:cartoons2}). Using this approach, the free energy barriers
 for a Gaussian chain\cite{sung96,muthu99,muthuentropy}, trapped initially inside a spherical cavity,  can be computed exactly and are purely 
entropic in nature due to lower degrees of conformational freedom in ``end-fixed'' state as compared with ``free ends'' and the 
absence of interactions. Similar calculations for the excluded volume chain\cite{muthu04} have been carried out within spherical 
symmetry and in the absence of solvent. For the case of polyelectrolyte translocation,
these barriers are unknown and form the focus of this study.

  Earlier theories of translocation of polyelectrolytes have been constructed only by using results of neutral polymer chains\cite{sung96,muthu99,muthuentropy} and ignoring the coupling between conformations of the polyelectrolyte chain and the small ions. 
However, it is widely being recognized that the physics of polyelectrolytes is dominated by counterions. In view of this, it becomes 
necessary to estimate the role of counterions and small electrolyte ions in establishing the free energy barriers for translocation. 
Here, we present a systematic calculation of free energy barrier by an explicit treatment of the coupling between small ions and conformations of polyelectrolyte chains. By adopting the self-consistent-field theory (SCFT) for a flexible polyelectrolyte chain and combining with the Poisson-Boltzmann prescription for the electrolyte ions and counterions, we have computed the various energetic and entropic contributions to the free energy barrier. Since the localization of one chain end at a specific location on the surface of the cavity breaks the radial symmetry, we have solved the self-consistent coupled nonlinear differential equations in two dimensions with azimuthal symmetry. As pointed out earlier, we address only the localization of one of the chain ends at the pore, without any consideration of all effects arising from the pore itself.

The rest of the paper is organized as follows: theory is presented in Sec. ~\ref{sec:theory}; numerical technique is presented in Sec. ~\ref{sec:numerics}; calculated results and conclusions are presented in Sec. ~\ref{sec:results} and ~\ref{sec:conclusions}, respectively.

\section{Theory}
\setcounter {equation} {0} \label{sec:theory}
In order to study confinement driven translocation, we consider a single negatively charged flexible polyelectrolyte chain 
in a spherical cavity of radius $R$ and model the chain as a continuous curve of length 
$Nb$, where $N$ is the number of Kuhn segments, each of length $b$. An arc length variable 
$t$ is used to represent any segment along the chain backbone so that $t \in [0,N]$. To maintain global electroneutrality, 
we assume that the spherical cavity is filled with $n_{c}$
monovalent counterions (positively charged) released by the chain in addition to
$n_{\gamma}$ ions of species $\gamma \,(= +,-)$ coming from added salt.
Moreover, we assume that there are $n_{s}$ solvent molecules (satisfying
the incompressibility constraint after assuming the small ions to be pointlike) present
in the cavity of volume $\Omega$ and for simplicity, each solvent molecule occupies a volume ($v_{s}$)
same as that of the monomer (i.e., $v_{s}\equiv b^{3}$).
Subscripts $p,s,c,+$ and $-$ are used to represent
monomers, solvent molecules, counterions from the polyelectrolyte, positive and negative salt ions, respectively.
The valency (with sign) of the charged species of type $j$ is represented by $Z_{j}$ and the 
degree of ionization of the chain is taken to be $\alpha$. In the point charge limit for the small ions considered here,
the cations from the added salt ($j=+$ ) and the counterions of the polyelectrolyte
chain ($j=c$) are indistinguishable from each other. Also, we consider smeared charge
distribution so that each of the segments carries a charge $e\alpha Z_{p}$, where $e$
is the electronic charge.

We use self-consistent field theory (SCFT) to compute the free energy of single flexible polyelectrolyte\cite{rajeev08} chain
in ``free ends'' and ``end-fixed'' states. Ignoring the potential interactions between solvent molecules and
small ions, the partition function for the single chain system in either of the states can be
written as
\begin{eqnarray}
       \exp\left(-\frac{F} {k_{B}T}\right )& = & \frac {1}{\prod_{j}n_{j}!}\int D[\mathbf{R}] \int \prod_{j} \prod_{m=1}^{n_{j}} d\mathbf{r}_{m} \quad \mbox{exp} \left \{-\frac {3}{2 b^2}\int_{0}^{N}
        dt\left(\frac{\partial \mathbf{R}(t)}{\partial t} \right )^{2} \right . \nonumber \\
&& -  \frac {1}{2} \int_{0}^{N}dt \int_{0}^{N}dt' V_{pp} [ \mathbf{R}(t) - \mathbf{R}(t')]
 -\sum_{j}\sum_{m =1 }^{n_{j}}  \int_{0}^{N}dt V_{pj} [ \mathbf{R}(t) - \mathbf{r}_{m}]
\nonumber \\
&& \left .      - \frac{1}{2}\sum_{j}\sum_{j'}\sum_{k =1 }^{n_{j} }\sum_{m =1 }^{n_{j'}} V_{jj'} [ \mathbf{r}_{k} - \mathbf{r}_{m}] \right \}\prod_{\mathbf{r}}\delta\left(\int_{0}^{N} dt \, \delta \left[\mathbf{r}-\mathbf{R}(t)\right] \right . \nonumber \\
     && \left . + \sum_{j=1}^{n_{s}} \delta \left[\mathbf{r}-\mathbf{r}_{j}\right] - \rho_{0}\right), \label{eq:parti_sing}
\end{eqnarray}
where $\mathbf{R}(t)$ represents the position vector for $t^{th}$ segment and subscripts $j,j' = s,c,+,-$. 
What distinguishes the ``end-fixed'' state from the ``free ends'' state in Eq. (~\ref{eq:parti_sing}) 
is the functional integral over $\mathbf{R}$. Physically, the functional integral over $\mathbf{R}$ represents the sum over all 
the possible conformations of the chain originating from one end and ending at the other. Explicitly, for the ``free ends'' state, the functional 
integral is given by $\int D[\mathbf{R}] \equiv \int d\mathbf{r}_0 \int d\mathbf{r}_N \int_{\mathbf{r}_0}^{\mathbf{r}_N} D[\mathbf{R}]$, 
where $\mathbf{r}_0$ and $\mathbf{r}_N$ are the positions of the ends of the chain represented by the specific values of the contour 
variable $t=0$ and $t=N$ along the chain, respectively. Similarly, $\int D[\mathbf{R}] \equiv \int d\mathbf{r}_N \int_{\mathbf{r}_0}^{\mathbf{r}_N} D[\mathbf{R}]$ for the chain, whose one end is fixed at $\mathbf{r}_0$ in the ``end-fixed'' state.

Note that in Eq. (~\ref{eq:parti_sing}), it is understood that the factor of $1/2$ in the last term
inside the exponent is present, only when $j=j'$ and $k_{B}T$ is the Boltzmann constant times absolute
temperature. Furthermore, $V_{pp}(\mathbf{r}),V_{ss}(\mathbf{r}) \mbox{ and}\: V_{ps}(\mathbf{r})$ represent
the interaction energies for monomer-monomer, solvent-solvent and monomer-solvent pairs, respectively,
when the interacting species are separated
by distance $r = \mid \mathbf{r} \mid $ and are given by
\begin{eqnarray}
       V_{pp}(\mathbf{r}) & = & w_{pp} \delta(\mathbf{r}) + \frac {Z_{p}^{2}e^{2}\alpha^{2}}{4\pi \epsilon_o \epsilon
       k_{B}T}\frac{1}{r}, \\
       V_{ss}(\mathbf{r}) & = & w_{ss} \delta(\mathbf{r}), \\
       V_{ps}(\mathbf{r}) & = & w_{ps} \delta(\mathbf{r}).
\end{eqnarray}

In writing the interaction energies, the short range excluded volume interactions are 
modeled by three dimensional delta functions $\left[\delta(\mathbf{r})\right]$ multiplied by the respective excluded volume parameters. For the monomer-monomer, 
solvent-solvent and monomer-solvent pairs, these parameters are taken to be $w_{pp}, w_{ss}$ and $w_{ps}$, respectively.  
Also, the long range electrostatic interactions are modeled by Coulomb's law after assuming the effective 
dielectric constant ($\epsilon$) of the medium to be position independent, $\epsilon_{o}$ being the permittivity of the vacuum. 

In writing Eq. (~\ref{eq:parti_sing}), the constraint on the number densities of 
the monomers and the solvent molecules to obey the incompressibility condition at all points inside the spherical cavity is written as the product of delta functions involving microscopic densities
on the right hand side ($\rho_{0}$
being the total number density of the system so that $\rho_{0} = (N + n_{s})/\Omega \equiv 1/b^{3}$ ). As mentioned earlier, the incompressibility condition 
is written after taking the small ions to be point-like. For the point-like limit of the small ions, the interaction energies between the monomers and the ions, represented by $V_{pj}$, are given by
       \begin{eqnarray}
       V_{pj}(\mathbf{r}) & = & \frac {Z_{p}Z_{j}e^{2}\alpha }{4\pi \epsilon_o \epsilon k_{B}T}\frac{1}{r}\quad \mbox{for} \quad j = c,+,-.
       \end{eqnarray}
Similarly, the interaction energies between the small ions can be written as

       \begin{eqnarray}
       V_{jj'}(\mathbf{r}) & = & \frac {Z_{j}Z_{j'}e^{2}}{4\pi \epsilon_o \epsilon k_{B}T}\frac{1}{r} \quad \mbox{for} \quad j,j' = c,+,-.
\end{eqnarray}

Following the protocol presented in Appendix A, the free energy $F$ of the single chain system in either of the 
states can be computed using the well-known saddle point approximation. 
$F$ is expressed as integrals over inhomogeneous number densities of the various components of the system and the electric potential. Taking the dielectric constant ($\epsilon$) of the medium to be independent of
temperature ($T$) and Flory's chi parameter defined as 
$ \chi_{ps}b^{3} =  w_{ps} - (w_{pp} + w_{ss})/2 \sim 1/T$, the free energy (within saddle-point approximation) can be divided\cite{marcus55}(see Appendix B and C for details) into
enthalpic part due to excluded volume and electrostatic interactions, and
entropic part due to small ions, solvent molecules and the polyelectrolyte chain.
Denoting these contributions by $E_{w},E_{e}, S_{ions}, S_{solvent}$ and  $S_{poly}$, respectively, the free energy
can be written as

\begin{eqnarray}
F^{\star} - F_{0} &=& E_{w} + E_{e} - T (S_{ions} + S_{solvent} + S_{poly}),  \label{eq:free_thermo_exp}
\end{eqnarray}
where $F_{0} = \frac{\rho_{0}}{2} \left( Nw_{pp} + n_{s}w_{ss} \right )$. Superscript $\star$ denotes that the saddle point approximation has
been used to compute the free energy. Explicit expressions for the different contributions are given by
\begin{eqnarray}
E_{w} &=& \chi_{ps}b^{3}\int d\mathbf{r}  \rho_{p}(\mathbf{r})\rho_{s}(\mathbf{r}), \label{eq:ew_exp}\\
E_{e} &=& \frac{1}{2} \int d\mathbf{r} \,\psi(\mathbf{r})\rho_{e}(\mathbf{r}), \label{eq:ee}\\
- T S_{ions} &=&  \sum_{j=c,+,-}\int d\mathbf{r}\,\rho_{j}(\mathbf{r})\left\{\ln\left[\rho_{j}(\mathbf{r})\right] - 1\right \},\label{eq:tsmall}\\
-T S_{solvent} &=&  \int d\mathbf{r}\,\rho_{s}(\mathbf{r})\left \{ \ln\left[\rho_{s}(\mathbf{r})\right]- 1 \right \},\label{eq:solvent_ent} \\
- T S_{poly} &=&  - \ln Q_{p} - \int d\mathbf{r}\, \left [ \left\{Z_{p}\alpha \psi(\mathbf{r}) + w_{p}(\mathbf{r}) \right \} \rho_{p}(\mathbf{r}) \right ]. \label{eq:poly_ent}
\end{eqnarray}
In these equations, $\rho_{\beta}(\mathbf{r})$ and $w_{\beta}(\mathbf{r})$ are respectively
the macroscopic number density and the field experienced by species of type $\beta$, due to excluded volume interactions 
\textit{at} the saddle point.
Also, all the charged species experience an electrostatic potential
represented by $\psi(\mathbf{r})$, which is related to the local charge density
$\rho_{e}(\mathbf{r}) = \sum_{j = c,+,-}Z_{j}\rho_{j}(\mathbf{r}) + Z_{p}\alpha \rho_{p}(\mathbf{r})$
by Poisson's equation, 
\begin{eqnarray}
\nabla_{\mathbf{r}}^{2}\psi(\mathbf{r}) &=& - 4\pi l_{B}\rho_{e}(\mathbf{r}). \label{eq:saddle_poisson}
\end{eqnarray}

Note that $\psi(\mathbf{r})$ in these equations is dimensionless
(in units of $k_{B}T/e$) and $l_{B}$ is the Bjerrum length defined as $l_{B} = e^{2}/4\pi\epsilon_0 \epsilon k_{B}T$.

At the saddle point, the macroscopic densities for the small molecules
are related to the corresponding fields by the
Boltzmann law, so that
\begin{eqnarray}
\rho_{s}(\mathbf{r}) &=& \frac{n_{s}\exp\left [-w_{s}(\mathbf{r}) \right]}{\int d\mathbf{r}\exp\left [-w_{s}(\mathbf{r}) \right]},\label{eq:saddle_exp1}\\
\rho_{j}(\mathbf{r}) &=& \frac{n_{j}\exp\left [- Z_{j}\psi(\mathbf{r}) \right]}{\int d\mathbf{r}\exp\left [-Z_{j}\psi(\mathbf{r}) \right]} \quad \mbox{for}\quad j = c,+,-. \label{eq:den_small}
\end{eqnarray}
The fields and densities are related to each other by the saddle point equations, given by
\begin{eqnarray}
w_{p}(\mathbf{r}) &=& \chi_{ps}b^{3}\rho_{s}(\mathbf{r}) + \eta(\mathbf{r}),\\
w_{s}(\mathbf{r}) &=& \chi_{ps}b^{3}\rho_{p}(\mathbf{r}) + \eta(\mathbf{r}),
\end{eqnarray}
where $\eta(\mathbf{r})$ is the
Lagrange's multiplier introduced to enforce the incompressibility
constraint. For the ``free ends'' state, the monomer density is dependent on the field by the relation
\begin{equation}
\rho_{p}(\mathbf{r}) \equiv \rho_{p}^{f}(\mathbf{r}) =  \frac{\int_{0}^{N} dt \, q(\mathbf{r},t) q(\mathbf{r},N-t)}{\int d\mathbf{r}q(\mathbf{r},N)} \label{eq:den_eq}
\end{equation}
and for the ``end-fixed'' state, the relation becomes
\begin{equation}
\rho_{p}(\mathbf{r}) \equiv \rho_{p}^{a}(\mathbf{r},\mathbf{r_{a}}) =  \frac{\int_{0}^{N} dt \, G(\mathbf{r},\mathbf{r_{a}},t,0) q(\mathbf{r},N-t)}{\int d\mathbf{r}G(\mathbf{r},\mathbf{r_{a}},N,0)}. \label{eq:den_eq_the}
\end{equation}
Superscript $f$ and $a$ depict the free and anchored nature of the single chain. These monomer densities are related to the solvent density by the incompressibility constraint
\begin{equation}
 \rho_{p}(\mathbf{r}) + \rho_{s}(\mathbf{r}) = \rho_{0} . \label{eq:saddle_incomp}
\end{equation}

In Eqs. (~\ref{eq:den_eq}) and (~\ref{eq:den_eq_the}), the function $q(\mathbf{r},t)$ is the probability of
finding segment $t$ at location $\mathbf{r}$, when starting end of the chain can be anywhere inside
the spherical cavity, and
it satisfies the modified diffusion equation
\begin{eqnarray}
\frac{\partial q(\mathbf{r},t) }{\partial t} &=& \left [\frac{b^{2}}{6}\nabla_{\mathbf{r}}^{2}- \left \{ Z_{p}\alpha \psi(\mathbf{r}) + w_{p}(\mathbf{r})\right \} \right ]q(\mathbf{r},t),  \label{eq:difreal}
\end{eqnarray}
along with the initial condition $q(\mathbf{r},0) = 1$. Similar to $q(\mathbf{r},t)$, the Green function, $G(\mathbf{r},\mathbf{r_{a}},t,0)$, is the probability of
finding segment $t$ at location $\mathbf{r}$, when starting end of the chain is at $\mathbf{r_{a}}$.
It also satisfies Eq. (~\ref{eq:difreal}) but with the initial condition $G(\mathbf{r},\mathbf{r_{a}},0,0) = \delta(\mathbf{r}-\mathbf{r_{a}})$, where $\delta$ represents the three dimensional delta function.
The partition function of the chain ($Q_{p}$) can be written in terms of these functions. Specifically,
for the ``free ends'' state, it is given by $Q_{p} \equiv  Q_{p}^{f} = \int d\mathbf{r}q(\mathbf{r},N)$
and for the ``end-fixed'' state, the partition function
becomes $Q_{p} \equiv Q_{p}^{a} = \int d\mathbf{r}G(\mathbf{r},\mathbf{r_{a}},N,0)$.

We include the effect of confinement by solving Eqs. (~\ref{eq:saddle_poisson} - ~\ref{eq:difreal}) using the Dirichlet boundary conditions for $w_s(\mathbf{r}),\psi(\mathbf{r})$, $q(\mathbf{r},t)$ and $G(\mathbf{r},\mathbf{r_{a}},t,0)$. Rationale for choosing these boundary 
conditions is that we model the spherical cavity as a neutral, hard surface. Dirichlet boundary conditions for  
$\psi(\mathbf{r})$ corresponds to a neutral spherical cavity. Similarly, Dirichlet boundary conditions for $q(\mathbf{r},t)$ and $G(\mathbf{r},\mathbf{r_{a}},t,0)$ means the 
spherical cavity is like a hard wall and the monomer density at the surface must be zero. 
This is equivalent to carrying out calculations with a wall potential, which is a delta function. The reason for choosing the Dirichlet boundary condition for $w_s(\mathbf{r})$ is the following: physically, the incompressibility constraint gets violated near the surface of the confining spherical cavity and due to the 
hard wall like spherical cavity, monomer as well solvent density at the surface 
must be zero. Masking techniques\cite{masking} have been used to take care of this fine 
point and study the surface effects. In this work, we take a different approach. 
Noticing that the free energy contribution from the values of densities and fields at the hard surface is zero, we choose $w_s(R) = 0$ and compute the solvent density at the surface. Note that pinning $w_s(R)$ to zero also means that the field is shifted by a constant. However, it is well-known that the densities and the free energy at the saddle-point are independent of this shift.  
   
For the ``free ends'' state, these equations 
have been solved after assuming radial symmetry\cite{rajeev08}. Since the radial symmetry is broken when one end of the chain is anchored near the surface at the site of the pore,  we have solved the SCFT equations in two dimensions with the azimuthal symmetry. In two dimensions, due to inherent 
singularities at the center of the spherical cavity (due to the division by zero) in addition to those arising from the delta functional form for 
the initial condition of the modified diffusion 
 equation for the ``end-fixed'' state, the numerical solution of SCFT equations is non-trivial. In the present work, we 
have developed an efficient numerical scheme to solve these coupled non-linear equations in the presence of aforementioned singularities, which is 
presented in the next section.

\section{Numerical Technique} \label{sec:numerics}
As mentioned earlier, we need to solve the above non-linear set of equations in two dimensions under the assumption of azimuthal symmetry. Note that the ``end-fixed'' state 
has an azimuthal symmetry in a polar coordinate system, where the origin is at the intersection of the surface of the confining sphere and the radius connecting the center of the sphere to the anchoring point. This means that in the coordinate system, where the origin is the center of the confining sphere (Fig. ~\ref{fig:cartoons2}c), the SCFT equations need to be solved on a semi-circle with its diameter along the x-axis. The solution on a semi-circle can be 
obtained by defining $\theta$ 
as the angle with respect to the only diameter of the semi-circle so that the SCFT equations need to be solved in $\left\{r,\theta\right\}$ space, where $r\in [0,R], \theta \in [0,\pi]$. In practice, this procedure can be carried out using $\nabla_{\mathbf{r}}^{2} = \frac{1}{r^{2}}\left[\frac{\partial}{\partial r}(r^{2}\frac{\partial}{\partial r}) + \frac{1}{\sin \theta}\frac{ \partial}{\partial \theta}\left(\sin \theta \frac{\partial}{\partial \theta}\right)\right]$.  This means that in order to use the pseudo-spectral method\cite{fleck88,saxena02} for an accurate solution of the modified diffusion equation, we need to use the Legendre polynomials and hence, the Legendre transform. This particular step slows down the 
computational procedure due to the unavailability of an efficient fast Legendre transform. 

In order to develop a faster algorithm, we note that for the particular value of $\theta = \pi/2$, the Laplacian becomes $\nabla_{\mathbf{r}}^{2} = \frac{1}{r^{2}}\left[\frac{\partial}{\partial r}(r^{2}\frac{\partial}{\partial r}) + \frac{\partial^{2}}{\partial \phi^{2}}\right]$. Due to the particular functional form of the $\phi$ dependent part in the Laplacian, fast Fourier transform\cite{fftw} (FFT) can be 
used to apply this part of the Laplacian. In other words, we need to solve on a circle  so 
that $r\in [0,R], \phi \in [0,2\pi]$ rather than a semi-circle. However, the use of 
FFT in the computations over the Legendre transform speeds up the calculations even when the number of collocation points on the grid gets enlarged by a factor of four. Also, note that we expect the results (for the densities etc. ) on a circle to be symmetric about the axis passing through the anchoring point and the center of the sphere due to the azimuthal symmetry of the problem. This provides a nice check on the computational procedure.   

In view of this computational efficiency, we have solved the SCFT equations using spherical polar co-ordinates ($r,\theta = \pi/2,\phi$) so that $r\in [0,R]$ and $\phi \in [0,2\pi]$. Instead of solving these equations in real or Fourier space, we use the split-step pseudo-spectral method\cite{fleck88,saxena02} employing the fast Fourier transform\cite{fftw} (FFT) and sine transform\cite{numericalrecipe}, which allows a faster and accurate computation of the densities and free energies. It is convenient to solve for $f(\mathbf{r},t) = rq(\mathbf{r},t)$ rather than solving for $q(\mathbf{r},t)$ directly. As earlier, the solution of Eq. (~\ref{eq:difreal}) is obtained after using $\nabla_{\mathbf{r}}^{2} = \frac{1}{r^{2}}\left[\frac{\partial}{\partial r}(r^{2}\frac{\partial}{\partial r}) + \frac{\partial^{2}}{\partial \phi^{2}}\right]$ and expanding $f(\mathbf{r},t) = \sum_{l=0}^{L}f_{l}(r,t)e^{il\phi}$, where $L$ is the number of terms required to represent the function $f$ by a finite series within the desired accuracy. Writing Eq. (~\ref{eq:difreal}) in terms of $f(\mathbf{r},t)$ and using Baker-Hausdorff formula\cite{fleck88,saxena02}, the solution is given by the propagation relation
\begin{eqnarray}
f(\mathbf{r},t + dt)  &\simeq&   \exp\left [-dt w(\mathbf{r})/2\right]\exp\left [dt\frac{b^{2}}{12}\frac{1}{r^{2}}\frac{\partial^{2}}{\partial \phi^{2}}\right] \nonumber \\
&&\exp\left[dt\frac{b^{2}}{6}\frac{\partial^{2}}{\partial r^{2}}\right]\exp\left [dt\frac{b^{2}}{12}\frac{1}{r^{2}}\frac{\partial^{2}}{\partial \phi^{2}}\right]\nonumber \\
&& \exp\left [-dt w(\mathbf{r})/2\right]f(\mathbf{r},t),\label{eq:split}
\end{eqnarray}
where $w(\mathbf{r}) = Z_{p}\alpha \psi(\mathbf{r}) + w_{p}(\mathbf{r})$. 
Also, the multiplication by $r$ leads to $f(0,\phi,t) = 0$ for all values of $t$ so that numerical problems at $r=0$, due to division by $0$ in the Laplacian are avoided. Another major advantage of the transformation is that now, the equations are to be solved with periodic boundary conditions. So, FFT and sine transform can be used to implement the exponential of operators. Exponential of $\phi$ dependent operator on the right hand side of Eq. (~\ref{eq:split}) is applied in Fourier space after taking one forward and one backward FFT. $r$ dependent operator is applied after taking forward sine transform defined as
\begin{eqnarray}
f_{l}(r,t)  &= &  \sum_{k=1}^{K}g_{k}(l,t)\sin(k\pi r/R),\label{eq:sine_transform}
\end{eqnarray}
and the corresponding inverse sine transform, so that boundary conditions for $f$ are always satisfied during the computations. Here, $K$ is the number of 
terms in the finite series on the right hand side of Eq. (~\ref{eq:sine_transform}) to accurately sample the function 
$f_l(r,t)$ within the desired accuracy. In total, each time step requires two FFTs with respect to $\phi$ and one sine transform with respect to $r$ in each direction (forward and backward). The same technique has been used for solving $G(\mathbf{r},\mathbf{r_{a}},t,0)$ 
after approximating the initial condition for $G(\mathbf{r},\mathbf{r_{a}},0,0) = \delta\left(\mathbf{r}-\mathbf{r_{a}}\right)$ by 
\begin{equation}
\delta\left(\mathbf{r}-\mathbf{r_{a}}\right) = \left\{\begin{array}{ll}
\frac{1}{2 r_a^2 \Delta \phi  \Delta r} & \quad \mbox{if}\quad \mathbf{r}=\mathbf{r_{a}} \\
0 & \quad \mbox{if}\quad \mathbf{r} \neq  \mathbf{r_{a}}.
\end{array}
\right .
\end{equation}
Here, $\Delta \phi = 2\pi/L$ and $\Delta r = R/K$ are the grid spacings used to discretize the two dimensional space. 

To solve Poisson's equation,$\nabla_{\mathbf{r}}^{2}\psi(\mathbf{r}) = - 4\pi l_{B}\rho_{e}(\mathbf{r})$, we use a similar strategy. We solve for $h(\mathbf{r}) = r\psi(\mathbf{r})$, so that Poisson's equation becomes
\begin{eqnarray}
\left[\frac{\partial^{2}}{\partial r^{2}} + \frac{1}{r^{2}}\frac{\partial^{2}}{\partial \phi^{2}}\right]h(\mathbf{r}) &=& - 4\pi l_{B}r \rho_{e}(\mathbf{r}). \label{eq:poisson}
\end{eqnarray}
Now, expanding $h(\mathbf{r}) = \sum_{l=0}^{L}h_{l}(r)e^{il\phi}$, the equation for components $h_{l}$
becomes
\begin{eqnarray}
\left[\frac{\partial^{2}}{\partial r^{2}} - \frac{l^{2}}{r^{2}}\right]h_{l}(r) &=& - 4\pi l_{B}\mbox{FFT}_{\phi}\left[r \rho_{e}(\mathbf{r})\right]. \label{eq:poisson_fourier}
\end{eqnarray}
Here, the subscript $\phi$ means, FFT is to be taken with respect to $\phi$. These sets of equations can be readily solved for the real and imaginary parts of $h_{l}(r)$ due to the
tridiagonal nature of the finite difference equation set obtained with the constraints $h_{l}(0) = h_{l}(R) = 0$. Now, taking backward FFT with respect to $\phi$, $h(\mathbf{r})$ is obtained.

Starting from an initial guess for fields, $w_{p}(\mathbf{r}),w_{s}(\mathbf{r})$ and $\psi(\mathbf{r})$,
new fields and densities are computed using the method described above with the 
boundary conditions mentioned in the previous section. Simple mixing\cite{numericalrecipe}
is used to
obtain the new guess and the iterative process is continued until the difference in newly computed and
the guessed fields is of the order of $10^{-7}$. Using the converged solution for
the fields and densities, free energies for the ``free ends''(i.e., $F^\star \equiv F_f^\star$) and ``end-fixed'' ($F^\star \equiv F_a^\star$) 
states are computed. For the computation of the free energy barriers for the chain end to find the pore, we compute the difference 
$\Delta F = F_a^\star - F_f^\star$. To analyze different contributions to the free energy barriers, we also compute the  
differences in the enthalpic and entropic contributions, given by 
$\Delta E_w = E_w(\mbox{``end-fixed''}) - E_w(\mbox{``free ends''}), \Delta E_e = E_e(\mbox{``end-fixed''}) - E_e(\mbox{``free ends''}), 
-T\Delta S_{ions} = -TS_{ions}(\mbox{``end-fixed''}) + TS_{ions}(\mbox{``free ends''}), -T\Delta S_{poly} = -TS_{poly}(\mbox{``end-fixed''}) + TS_{poly}(\mbox{``free ends''})$ and $-T\Delta S_{solvent} = -TS_{solvent}(\mbox{``end-fixed''}) + TS_{solvent}(\mbox{``free ends''})$. 
The results presented here were obtained by using
$L=32, K=32$ and $dt = 0.1$ after optimizing the numerical algorithm for speed and accuracy.
Also, we have fixed one end of the chain for ``end-fixed'' state at $\mathbf{r_{a}} = \left[(R-0.625)b,\pi/2,0\right]$ in spherical polar co-ordinates . 

\section{Results} \label{sec:results}
In Sec. ~\ref{sec:theory}, we have presented the theoretical analysis for a flexible polyelectrolyte chain in the 
presence of counterions and coions of arbitray valency. However, it is well-known in the literature that the saddle-point approximation 
used in this work (which provides the Poisson-Boltzmann description for the electrolytes) fails\cite{orland97,vlachy99,netz2000} in the case of 
multivalent ions. For monovalent ions\cite{vlachy99}, Poisson-Boltzmann equation provides quite reasonable results. In view of this, we have considered here  
a polyelectrolyte chain with monovalent counterions in the presence of monovalent salt. 

For monovalent counterions and coions, we have carried out 
extensive numerical computations of the free energy barriers by varying the various parameters required to solve the above coupled equations 
namely, $N$, $R$, $\alpha$, $l_B$, $\chi_{ps}$, and the salt concentration $c_s$ (in units of moles per liter, i.e., $M$). For wide ranges of $\alpha$, $l_B$, $\chi_{ps}$, and $c_s$, we have solved the above equations for the $N-$ and $R-$ dependencies of the monomer, counterion and coion density profiles, 
electric potential distribution and the various free energy contributions. It turns out that these extensive numerical calculations lead to some 
general conclusions, presented below. 

\subsection{Monomer, counterion and coion distributions}
Typical monomer and electrostatic potential distributions for the ``free ends'' and ``end-fixed'' states are shown in Fig. ~\ref{fig:den_pot_n50} 
for $N=50$ and $R/b=4$. Here, we have chosen $\alpha = 0.1, l_B = 3 b, \chi_{ps} = 0.45$, 
and $c_s = 0.1 M$, where the choice of the parameters has been motivated by the experimental relevance to aqueous systems. Although these parameters cannot be varied independently in experimental situations, we have computed the consequences of each of these parameters in order to obtain physical insight into the origins of the free energy barriers. It is evident from Figs. ~\ref{fig:den_pot_n50}a 
and ~\ref{fig:den_pot_n50}b that, for the unanchored state, that both distributions are radially symmetric, with the monomer density and the negative electric potential being a maximum at the center of the cavity. 
As one end of the chain is localized near the right-edge of the equator (represented by 
the arrow in Figs. ~\ref{fig:den_pot_n50}c and ~\ref{fig:den_pot_n50}d), the density and potential distributions become anisotropic as expected. 

Also, the coupling between the counterion, coion and monomer distributions, whose origin lies in the electrostatic 
interactions between these charged species, can be seen clearly in Fig. ~\ref{fig:ions_n50}. In the figure, we have plotted the counterion and 
coion distributions for the same single chain systems, whose monomer and potential distributions are shown in Fig. ~\ref{fig:den_pot_n50}.  
For the ``free ends'' state, it is found that the counterion and coion distributions (Figs. ~\ref{fig:ions_n50}a 
and ~\ref{fig:ions_n50}b, respectively) are also radially symmetric like the monomer distribution shown in 
Fig. ~\ref{fig:ions_n50}a. Furthermore, Figs. ~\ref{fig:den_pot_n50}a and ~\ref{fig:ions_n50}a show that the counterion distribution 
tracks the monomer density distribution with the maximum number density of the counterions at the center of the cavity. 
On the other hand, Fig. ~\ref{fig:ions_n50}b shows that the coions are excluded from the 
regions rich in monomer density with a minimum number density of
the coions at the center of the cavity. These results are consistent with the calculations carried out for the ``free ends'' state with 
radial symmetry\cite{rajeev08}. However, Figs. ~\ref{fig:ions_n50}c and ~\ref{fig:ions_n50}d show that the 
radial symmetry gets broken for the ``end-fixed'' state and the anisotropy of the small ion distributions follow 
from the electrostatic coupling between counterion, coion 
and monomer distributions. In addition, it is to be noted that although the counterions and coions are distributed differently for the ``free ends'' 
and ``end-fixed'' states, the net electric potential ( Figs. ~\ref{fig:den_pot_n50}b and ~\ref{fig:den_pot_n50}d) tracks the monomer density distribution.

These results (Figs. ~\ref{fig:den_pot_n50} and ~\ref{fig:ions_n50}) show that the anchoring of the chain end at a specified location 
leads to anisotropic monomer and charge distributions. 
Intuitively, we anticipate that larger anisotropies in the monomer and charge distributions for the ``end-fixed'' state relative to 
the ``free ends'' state correspond to larger free energy differences 
between the states and hence, larger free energy barriers. Also, due to the coupling between the monomer 
and small ion distributions, a longer chain in ``end-fixed'' state is expected to show a higher degree of anisotropy in the monomer and charge distributions 
compared to a shorter chain. However, in a confined space, the increase in the chain length leads to space filling and counteracts the 
anisotropic distributions, which is otherwise expected. We demonstrate this particular point in Fig. ~\ref{fig:den_pot_n100} 
by presenting the monomer and electrostatic potential distribution for the chain with twice the number of monomers as in Fig. ~\ref{fig:den_pot_n50}
(i.e., $N=100$) and keeping all other parameters the same.  

Comparing Figs. ~\ref{fig:den_pot_n50}a with ~\ref{fig:den_pot_n100}a, and ~\ref{fig:den_pot_n50}c with ~\ref{fig:den_pot_n100}c, 
it is clear that the increase in the chain length for a fixed radius of the confining spherical cavity leads to an increase in 
the monomer density at all points in the interior for both the ``free ends'' and ``end-fixed'' states as expected. 
Furthermore, monomer density distributions for ``free ends'' and ``end-fixed'' states (Figs. ~\ref{fig:den_pot_n100}a 
and ~\ref{fig:den_pot_n100}c, respectively) are found to be almost indistinguishable from each other except a small region near 
the anchoring point. Similar trends are seen for the electrostatic potential (compare Fig. ~\ref{fig:den_pot_n100}b with ~\ref{fig:den_pot_n100}d) 
and small ion distributions (not shown). 

To be more quantitative, we have plotted the monomer density profiles in the 
``free ends'' and ``end-fixed'' states along $x$ and $y$ axes for different values of $N$ for a given spherical cavity (Fig. ~\ref{fig:den_all_n}). We have plotted these 
density profiles for the values of $N$ at which we see noticeable differences in the 
free energy barriers for the polyelectrolyte and the self-avoiding walk chains. 
From Figs. ~\ref{fig:den_all_n}a 
and ~\ref{fig:den_all_n}b, it is clear that the system has radial symmetry in ``free ends'' 
state. However, the radial symmetry is broken in ``end-fixed'' state as it is clear from 
Figs. ~\ref{fig:den_all_n}c and ~\ref{fig:den_all_n}d. Also, as mentioned earlier the monomer density 
increases everywhere inside the spherical cavity, when $N$ is increased for both the ``free-ends'' and ``end-fixed'' states. 

Furthermore, the expected increase in the degree of anisotropy for the monomer and charge distributions in the ``end-fixed'' state with the increase in the chain length does not occur for all monomer volume fractions in a confined system, where space filling 
effects counteract the anisotropy. This particular point is demonstrated in Fig. ~\ref{fig:densities_x_y_n}, where we have plotted the monomer density distributions along $x$ and $y$ axes in the ``free ends'' and ``end-fixed'' states at different monomer volume fractions. At a low volume fraction (corresponding to $N=50$ at $R/b=4$ in Figs. ~\ref{fig:densities_x_y_n}a and ~\ref{fig:densities_x_y_n}b), there is significant anisotropy in the monomer density distributions arising from the fixing of one end 
of the chain. However, at a higher volume fraction (corresponding to $N=100$ at $R/b=4$ in Figs. ~\ref{fig:densities_x_y_n}c and ~\ref{fig:densities_x_y_n}d), the monomer density 
distributions in the ``free ends'' and ``end-fixed'' states become almost indistinguishable from each other. Note that the electrostatic potential and the density distribution of the small ions also become indistinguishable from each other in the two states at higher volume fractions (Figs. ~\ref{fig:ions_x_n}a and ~\ref{fig:ions_x_n}b, respectively). This is a result of the space filling or confinement effect counteracting the anisotropic effects originating due to the localization of one of the ends of the chain. This particular point will be used later to explain the trends in the free energy differences in the ``free ends'' and ``end-fixed'' states presented below.

\subsection{Free energy barriers}
 In Fig. ~\ref{fig:barriers_r}, we have plotted the free energy difference (in units of $k_BT$) between the ``end-fixed''
 and ``free ends'' states for different values of $N$ and $R$. For comparison purposes, the barriers for the
 chains when electrostatics is switched off (i.e., self-avoiding walk chain with $\alpha=0$, $c_s=0$, $l_B=0$), are also plotted. 
It is evident that the free energy barriers for polyelectrolyte chains are almost identical to those for the corresponding uncharged self-avoiding 
walk chain at higher monomer densities and they differ only by a small amount in lower density regime. 
As seen in Fig. ~\ref{fig:barriers_r}, the dependence of the free energy barrier on the chain length is nonmonotonic, for a 
given cavity size. 

The nonmonotonicity in the free energy barriers has also been seen in the case of 
self-avoiding walk chains in the absence of solvent\cite{muthu04}. In agreement with  
Ref. \cite{muthu04}, the origin of the nonmonotonicity lies in the entropic, excluded volume and the confinement effects. For low enough values of $N$ such that the net intra-chain excluded volume interaction is weak, the probability\cite{muthuentropy} of finding a particular monomer (such as the chain end) at any prescribed spatial location decreases with an increase in $N$. This is equivalent  to an increase in the free energy barrier with $N$ in this limit. The entropic contributions to the free energy barriers arising from the anisotropic small ion distribution add to this effect (see the description below on the effect of electrostatics in Figs. ~\ref{fig:alpha_effect},~\ref{fig:cs_effect} and ~\ref{fig:lb_effect}). On the other hand, for higher packing fractions of monomers, the excluded volume and the synergistic space filling effects coming from the confinement take over and consequently the two equilibrium states become indistinguishable from each other (compare Figs. ~\ref{fig:den_pot_n50} and ~\ref{fig:den_pot_n100}). This is equivalent to a decrease in the free energy barrier with an increase 
in $N$  for the limit of strong confinement.  

Although the numerical values of $F_f^\star$ and $F_a^\star$ are significantly different for the polyelectrolyte  and uncharged polymer 
cases, the agreement between the free energy barriers for these two cases, as seen in
Fig. ~\ref{fig:barriers_r}, is striking. In order to identify the origins of the barriers for polyelectrolytes and of the agreement with 
uncharged polymers, we have plotted different energetic and
entropic contributions for two radii in Fig. ~\ref{fig:barrier_comps}. It is found that the dominant contribution to
the barriers is the difference in conformational entropy ($-T\Delta S_{poly}$) of the chain in the ``free ends'' and ``end-fixed'' states. 
Difference in solvent entropy ($-T\Delta S_{solvent}$) and energy ($\Delta E_w$) due to excluded volume interactions between different constituents 
plays a role, although meager, only when chain length is small and solvent is the major component in the system. Contributions due to the 
difference in entropy ($-T\Delta S_{ions}$) of small ions and electrostatic energy  ($\Delta E_e$) are negligible in comparison with 
other contributions. 

It has already been shown that the excluded volumer interaction energy ($E_w$), electrostatic energy ($E_e$) and 
conformational entropy ($-TS_{poly}$) of a flexible polyelectrolyte chain are minor 
contributors to the absolute chain free energy\cite{rajeev08}. Only for very strong confinements (with monomer volume fractions higher 
than $0.7$), the chain conformational entropy starts contributing significantly. Otherwise, it is the entropy of the 
small ions ($-TS_{ions}$) and the solvent molecules ($-TS_{solvent}$), which dominate the free energy. 
We have confirmed the same results using the two dimensional algorithm presented in Sec. ~\ref{sec:numerics} (comparison not shown here). 
Based on the relative contributions to the total free energy coming from different components in the system, one might think that the change 
in counterion, coion or solvent distribution would dominate the free energy barriers. However, on the contrary, we have found that the dominant 
contribution to the free energy barriers is the change in the conformational entropy of the chain. 

Analysis of the different contributions to the free energies, calculated for the different values of the parameters, leads to the following explanation. Entropies of small ions ($-TS_{ions}$) 
and the solvent molecules ($-TS_{solvent}$)  are strongly dependent on the number of 
these molecules in the system and show a very weak dependence on the spatial distributions (cf. Eqs. (~\ref{eq:tsmall}) and (~\ref{eq:solvent_ent})). 
For a single chain system in the presence of 
salt, a large number of small ions and solvent molecules explain the dominance of the total free energy by their entropies. 
However, due to an equal number of small ions and  solvent molecules in the ``free ends'' and ``end-fixed'' states, the entropies of the 
small ions and the solvent molecules are almost the 
same in either of the states and cancel each other almost exactly in the computation of the free energy barriers. This explains the minor contribution to the free energy barriers due to the small ions and solvent molecules in Fig. ~\ref{fig:barrier_comps}. 

Similarly, from Eq. (~\ref{eq:ew_exp}), the excluded volume interaction energy ($E_w$) depends on the monomer distribution. However, 
Figs. ~\ref{fig:den_pot_n50}a, ~\ref{fig:den_pot_n50}c, ~\ref{fig:den_pot_n100}a 
and ~\ref{fig:den_pot_n100}c reveal that an increase in the chain length for a given radius of the cavity makes 
the ``end-fixed'' state almost indistinguishable from the ``free ends'' state in terms of monomer distribution due to confinement effects. 
Consequently the excluded volume interaction energy contributions to the free energy barriers are very small at higher monomer volume fractions. Numerical results also show that the electrostatic energy ($E_e$) contribution to the total free energy is orders of magnitude lower than the other contributions (e.g., $E_e$ is of the order of $0.1$ for $F_f^\star =  -400$ for $R/b = 4$ in Fig. ~\ref{fig:barriers_r}). Keeping in mind such a low contribution to the total free energy, the negligible contribution to the 
free energy barriers ($\Delta F$) from the change in electrostatic energy ($\Delta E_e$) is not a surprise.  
However, the chain conformational entropy depends on the distribution of the chain ends (cf. Eq. ~\ref{eq:poly_ent}) 
and indeed differ in the two states due to a relatively lower number of conformational states available for 
the chain in the ``end-fixed'' state compared to the ``free ends'' state. As a result, the difference ($-T\Delta S_{poly}$) shows up as the dominant contribution 
to the free energy barriers ($\Delta F$) in Fig. ~\ref{fig:barrier_comps}. Furthermore, comparing 
Figs. ~\ref{fig:barrier_comps}a and ~\ref{fig:barrier_comps}b, it is found that in addition to the dependence 
of the barriers directly on the monomer density, the cavity radius plays an important role in modifying the barriers.  In agreement with the work on Gaussian chains\cite{muthuentropy}, the cavity radius provides an entropic contribution to the barriers going like $4\ln (R/l)$. This explains the decrease in the free energy barriers with a decrease in $R$, as is evident in Fig. ~\ref{fig:barriers_r}.

These results indeed support the entropic nature of the free energy barriers. Furthermore,
these results are quite robust (within $1-2 \, k_B T$) for a vast range of parameters involving $\alpha, l_B/b, \chi_{ps}$ and $c_s$. In Fig. ~\ref{fig:alpha_effect}, we 
have plotted the free energy barriers for different values of the degree of ionization ($\alpha$) of the polyelectrolyte chain. It is found that indeed electrostatics play a role in the free energy barriers and that the 
barriers increase with the increase in the degree of ionization of the chain at lower volume fractions. 
However, the increase in the free energy barriers is small (e.g., the change in the free energy barriers is less than $10 \%$ when $\alpha$ is changed from $0.1$ to $0.5$ in Fig. ~\ref{fig:alpha_effect}). 
Reason for the small change in the 
absolute value of the free energy barriers is the already mentioned logarithmic dependence of the small ion entropy 
term (i.e., $-T\Delta S_{ions}$) on the anisotropic ion distributions (cf. Eq. ~\ref{eq:tsmall}) and a weak dependence 
of the chain entropy (i.e.,$-T\Delta S_{poly}$) on $\alpha$.
Although the \textit{absolute} change in the free energy barriers is small when the degree of ionization is changed by 
a factor of $5$, it is worthwhile to investigate the origin of the change in the free energy barriers. It is found that 
the change in the free energy barriers has a dominant (but small in magnitude) contribution coming from the small 
ion entropy (e.g., $-T\Delta S_{ions}$ changes from $0.03$ to $0.44$ compared to the 
change of $-T\Delta S_{poly}$ from $6.11$ to $6.28$ when $\alpha$ is changed from $0.1$ to $0.5$ in 
Fig. ~\ref{fig:alpha_effect} at the lowest volume fraction). In other words, the slight increase in the 
free energy barriers comes mainly from the anisotropic distribution of the small ions in the ``end-fixed'' state. At higher monomer volume fractions, the confinement effects take over and the barriers are the 
same as for the self-avoiding walks. Similarly, the effect of added salt concentration and Bjerrum length on the free energy barriers are shown in Fig. ~\ref{fig:cs_effect} and ~\ref{fig:lb_effect}, respectively. Similar to the effect of $\alpha$, the free energy barriers increase with the lowering of the salt concentration and an increase in Bjerrum length i.e., with the strengthening of the electrostatics. Note that the change in the free energy barriers due to the change in 
Bjerrum length is minuscule due to the weak contribution of the electrostatic energy (cf. Eq. ~\ref{eq:ee}) 
to the free energies\cite{rajeev08}. 

The effect of solvent quality on the free energy barriers is shown in Fig. ~\ref{fig:chi_effect}. From the figure, it is clear that an increase in the solvent quality i.e., a decrease in $\chi_{ps}$, leads to an increase in the free energy barriers. This is an outcome of the change in excluded volume interaction energy with the change in 
the solvent quality (see Eq. ~\ref{eq:ew_exp}) and can be explained as following. 
With the change in the solvent quality, it is found that all the contributions to the free energy barriers 
remain almost the same except the excluded volume interaction energy (i.e., $\Delta E_w$ given by Eq. ~\ref{eq:ew_exp}). 
Note that the excluded volume interaction energy  
depends quadratically on the monomer density distribution 
(because $\chi_{ps}b^3\int d\mathbf{r}\rho_p(\mathbf{r})\rho_s(\mathbf{r}) = \chi_{ps}b^3N - \chi_{ps}b^3\int d\mathbf{r}\rho_p^2(\mathbf{r})$ as a result of the incompressibility constraint). Due to the higher monomer density near the surface 
of the confining cavity in the ``end-fixed'' state compared to the ``free ends'' state, $\Delta E_w$ is negative 
(cf. Fig. ~\ref{fig:barrier_comps}). Also, the prefactor $\chi_{ps}$ in Eq. ~\ref{eq:ew_exp} causes a decrease in the 
magnitude of $\Delta E_w$ (which is \textit{negative}) with the decrease in $\chi_{ps}$ (i.e., the increase in the solvent quality). In view of this, the free energy barriers increase with the increase in the solvent quality.     

\section{Conclusions}\label{sec:conclusions}
In summary, we find the remarkable result that even though the conformational entropy of a 
flexible polyelectrolyte chain is a minor contributor to the chain free energy, the free energy barrier is 
essentially entirely due to the change in the conformational entropy of the chain, for experimentally relevant 
conditions of translocation experiments. Even more remarkably, the free energy barrier for a flexible polyelectrolyte 
for moderate salt concentrations is not significantly different from that for an uncharged self-avoiding chain. The free 
energy barrier increases with degree of ionization, Bjerrum length, and solvent quality, and decreases with salt concentration. However, the increase in the free energy barriers in the confined single chain system investigated here is small. Furthermore, it is to be noted that the entropically driven free energy barrier for placing one chain end at the pore entrance is about $6-9 \, k_B T$, which is within the access of energy released in one event of ATP hydrolysis\cite{cellbook}. Nevertheless, in experiments involving fast translocations, the barriers 
computed here with equilibrium consideration might be modified by non-equilibrium polymer conformations.

Finally, it must be remarked that the present development of the model and numerical scheme to treat the anisotropy of polyelectrolyte conformations is only a starting point. The influences of electrostatic forces arising  from dielectric mismatches due to the pore-bearing membrane, and the nature of the charged pore itself are some of the future directions for integrating theories of polyelectrolyte translocation into experimental investigations.

\section*{ACKNOWLEDGEMENT}
The research was supported by the NIH Grant No. R01HG002776,
NSF Grant No. DMR $0706454$ and the MRSEC at the University of Massachusetts, Amherst.

\renewcommand{\theequation}{A-\arabic{equation}}
  \setcounter{equation}{0}  
  \section*{APPENDIX A : Self Consistent Field Theory }
Here, we present the details about the calculation of the free energy of a single polyelectrolyte chain within saddle point 
approximation. Similar procedure has been used earlier\cite{rajeev08,shi,wang,freedreview,fredbook}. In order to carry out the 
transformation from a description involving particles to the fields, we define 
a dimensionless Flory's chi parameter as $ \chi_{ps}b^{3} = w_{ps} - (w_{pp} + w_{ss})/2$ along with microscopic densities as
 \begin{eqnarray}
    \hat{\rho}_{p}(\mathbf{r})  &=& \int_{0}^{N} dt  \delta \left[\mathbf{r}-\mathbf{R}(t)\right], \\
     \hat{\rho}_{j}(\mathbf{r}) &=&  \sum_{i=1}^{n_{j}} \delta (\mathbf{r}-\mathbf{r}_{i}) \quad \mbox{for} \quad j = s,c,+,-,\\
\hat{\rho}_{e}(\mathbf{r}) &=& \alpha Z_{p} \hat{\rho}_{p}(\mathbf{r}) + \sum_{j = c,+,-}Z_{j}\hat{\rho}_{j}(\mathbf{r}) ,
\end{eqnarray}
where $\hat{\rho}_{p}(\mathbf{r}), \hat{\rho}_{j}(\mathbf{r})$ and $e\hat{\rho}_{e}(\mathbf{r})$
stand for monomer, small molecules (ions and solvent molecules) and local charge density, respectively (in units of $e$, $e$ being the charge of an electron).

Using these definitions, the partition function in Eq. (~\ref{eq:parti_sing}) can be rewritten as 
\begin{eqnarray}
       \exp\left(-\frac{F - F_0} {k_{B}T}\right )& = & \frac {1}{\prod_{j}n_{j}!}\int D[\mathbf{R}] \int \prod_{j} \prod_{m=1}^{n_{j}} d\mathbf{r}_{m} \quad \exp \left \{-\frac {3}{2 b^2}\int_{0}^{N}dt\left(\frac{\partial \mathbf{R}(t)}{\partial t} \right )^{2}
\right . \nonumber \\
&& \left . - \chi_{ps}b^{3}\int d\mathbf{r} \hat{\rho}_{p}(\mathbf{r})\hat{\rho}_{s}(\mathbf{r}) - \frac{l_B}{2}\int d\mathbf{r}\int d\mathbf{r}\,'\frac{\hat{\rho}_{e}(\mathbf{r})\hat{\rho}_{e}(\mathbf{r}\,')}{ |\mathbf{r}-\mathbf{r}\,'|}  \right \}\nonumber \\
&& \prod_{\mathbf{r}} \delta\left(\hat{\rho}_{p}(\mathbf{r}) + \hat{\rho}_{s}(\mathbf{r}) - \rho_{0}  \right), \label{eq:parti_den}
\end{eqnarray}
where $F_0/k_{B}T = (\rho_{0}/2) \left( Nw_{pp} + n_{s}w_{ss}\right )$ and $l_B = e^2/4\pi \epsilon_0 \epsilon k_B T$. 

To carry out the transformation from particles to fields, we use the following three transformations in order: (1) the well-known
Hubbard-Stratonovich transformation for the electrostatics part, which leads to the introduction of field $\psi$ in the calculations by
\begin{eqnarray}
\exp\left(-\frac{l_B}{2}\int d\mathbf{r} \int d\mathbf{r}\,'\frac{\hat{\rho}_{e}(\mathbf{r})\hat{\rho}_{e}(\mathbf{r}\,')}{|\mathbf{r}-\mathbf{r}\,'|}\right  ) &=& \frac{1}{\mu_{\psi}}\int
D[\psi(\mathbf{r})]\exp\left [ - \int d\mathbf{r} \left \{ i \psi(\mathbf{r})\hat{\rho}_{e}(\mathbf{r}) \right . \right. \nonumber \\
&& \left . \left .\qquad \qquad \qquad \qquad
- \frac{\psi(\mathbf{r})}{8 \pi l_{B}}\nabla_{\mathbf{r}}^{2}\psi(\mathbf{r})\right \}
\right ],
\end{eqnarray}
where $i = \sqrt{-1}$ is purely imaginary number and 
\begin{eqnarray}
\mu_{\psi} &=& \int D[\psi(\mathbf{r})]\exp\left [ \frac{1}{8 \pi l_{B}}\int d\mathbf{r}\psi(\mathbf{r})\nabla_{\mathbf{r}}^{2}\psi(\mathbf{r})                                       \right ].
\end{eqnarray}

(2) functional integral representation for unity to decouple the excluded volume interactions between the monomers and solvent molecules  
\begin{eqnarray}
1 &=& \int \prod_{k=p,s}D[w_k] \int  \prod_{k=p,s} D[\rho_k] \exp\left [ i \int d\mathbf{r} \sum_{k=p,s}w_{k}(\mathbf{r})\left \{ \rho_{k}(\mathbf{r}) - \hat{\rho}_{k}(\mathbf{r}) \right \} \right ], 
\end{eqnarray}
which leads to the introduction of collective fields $w_k$ and densities $\rho_k$ for the monomer and solvent molecules. 
Also, the transformation leads to the replacement of microscopic density variables ($\hat{\rho}_k$) by the collective density 
variables ($\rho_k$). 

(3) functional integral representation for the
delta functions to enforce incompressibility constraint at all points in the system  
\begin{eqnarray}
\prod_{r} \delta\left(\rho_{p}(\mathbf{r}) + \rho_{s}(\mathbf{r}) - \rho_{0}  \right)
 &=& \int D[\eta(\mathbf{r})]\exp\left[-i\int d\mathbf{r} \eta(\mathbf{r})\left(\rho_{p}(\mathbf{r}) +
\rho_{s}(\mathbf{r}) - \rho_{0}        \right) \right ],
 \end{eqnarray}
which introduces the well-known pressure field $\eta(\mathbf{r})$ in the calculations.

Using these transformations along with Stirling's approximation $\ln n! \simeq n \ln n - n$, Eq. (~\ref{eq:parti_den}) becomes
\begin{eqnarray}
 \exp\left(-\frac{F - F_0} {k_{B}T}\right )& = & \frac{1}{\mu_{\psi}}
\int\prod_{k=p,s} D[w_{k}] \int\prod_{k=p,s} D[\rho_{k}] \int D[\eta]\int D[\psi]
\exp\left [ - f\left\{w_{k},\rho_k,\eta,\psi\right \} \right ], \nonumber \\
&& \label{eq:functional_tobe}
\end{eqnarray}
where
\begin{eqnarray}
      f\left\{w_{k},\rho_{k},\eta,\psi\right \} & = & \chi_{ps}b^{3}\int d\mathbf{r} \rho_{p}(\mathbf{r}) \rho_{s}(\mathbf{r}) + i \int d\mathbf{r} \eta(\mathbf{r}) \left(\rho_{p}(\mathbf{r}) + \rho_{s}(\mathbf{r}) - \rho_{0} \right) - \ln Q_{p} \nonumber \\
&& - i \int d\mathbf{r} \sum_{k=p,s}w_{k}(\mathbf{r}) \rho_{k}(\mathbf{r}) + \sum_{j=s,c,+,-}n_{j}\left[\frac{n_j}{\ln Q_{j}} -1 \right] \nonumber \\ 
&& - \frac{1}{8\pi l_{B}}\int d\mathbf{r}\psi(\mathbf{r})\nabla_{\mathbf{r}}^{2}\psi(\mathbf{r}), \label{eq:parti_functional}
\end{eqnarray} 
where $Q_p,Q_s$ and $Q_j$ are the single particle paritition functions for the polyelectrolyte chain, solvent molecule and small ions of different species. 
Explicitly, the chain partition function is given by
\begin{eqnarray}
       Q_{p} & = & \int D[\mathbf{R}(t)]\exp \left [ -\int_{0}^{N}dt\left \{\frac{3}{2b^2}\left (\frac{\partial \mathbf{R}}{\partial t} \right )^{2}
+  i w_{p}\{\mathbf{R}\} + i Z_{p}\alpha \psi\{\mathbf{R}\} \right \} \right ].  
\end{eqnarray}
Similarly, single particle partition function for a solvent and small ion is given by
\begin{eqnarray}
Q_{s} & = & \int d\mathbf{r}\quad \mbox{exp} \left [ - iw_{s}(\mathbf{r})\right ], \\
Q_{j} & = & \int d\mathbf{r}\quad \mbox{exp} \left [ - iZ_{j}\psi(\mathbf{r})\right ] \quad \mbox{for} \quad j = c,+,-.
\end{eqnarray}

The functional integrations over the collective fields and densities in Eq. (~\ref{eq:functional_tobe}) are almost impossible to compute 
exactly. A well-known approximation to evaluate the functional integrals is the steepest descent technique
(also known as saddle-point approximation). We use the approximation to compute the 
free energy of the single chain system so that the approximated free energy is given by
$F -F_0 \simeq F^\star - F_0 = f\left\{w_{k}^{\star},\rho_{k}^{\star},\eta^\star,\psi^{\star}\right\}$ (after taking $k_BT = 1$ as a scale for energy), 
where collective fields and densities with stars as superscripts are their respective values at the saddle point and lead to the 
extremum of the functional $f$. Carrying out extremization of the functional $f$ with respect to the collective fields and densities, 
Eqs. (~\ref{eq:saddle_poisson}-~\ref{eq:saddle_incomp}) are obtained and details of carrying out the standard functional derivatives\cite{freedreview,fredbook}. Note that the normalization factor $\mu_\psi$ is ignored in the calculations for the approximate 
free energy by the saddle-point approximation.

\renewcommand{\theequation}{B-\arabic{equation}}
  \setcounter{equation}{0}  
  \section*{APPENDIX B : Energy and entropy of single polyelectrolyte chain within saddle point approximation}
 
Here, we present the calculations of the energetic and entropic contributions to the free energy
of a single polyelectrolytic chain using thermodynamic arguments. The analysis is similar to 
the one presented in Ref. \cite{marcus55}, where polymer conformations and hence, the monomer density 
are kept fixed during the well-known Debye charging process. For the free energy within the saddle point 
approximation used in this work, polymer conformations and hence, the monomer density are also dependent on the 
charging parameter and need to be treated properly. 

Computation of the energetic and entropic contributions to the final free energy within the saddle point 
approximation can be carried out after imagining two \textit{isothermal} ``charging'' processes. One is the 
traditional electric charging process, where charges of all the ions in the system
(free ions in the solution as well as on the chain backbone) are increased gradually from $0$ to their final values. 
 Similarly, we imagine  another ``charging'' process, where  excluded volume parameter characterizing the 
short range excluded volume effects is developed (or ``charged'') gradually from $0$ to its final value, 
which, in turn, leads to the development of monomer density and field waves.

For the electric charging, we assume that at any instance, charge of the ions is a 
fraction $\lambda$ of its final value. Similarly, the excluded volume ``charging'' process is 
characterized by the charging parameter $\lambda$, so that at any instance the excluded volume 
parameter characterizing the excluded volume interactions between species $i$ and $j$ is $w_{ij}' = \lambda^2 w_{ij}$, 
where $w_{ij}$ is the final value of the excluded volume parameter. Here, we follow the notation used by Marcus\cite{marcus55} to represent quantities at any instance in the charging process by superscript $'$ and the choice of 
$\lambda^2$ to relate the instant value of the excluded volume parameter to the final value is made to 
simplify the mathematics as discussed at the appropriate places in this Appendix. Also, the order of charging for the 
two processes does not matter as expected.    

Before computing the energy and entropy for the single polyelectrolyte chain with the solvent treated by using the incompressibility 
condition, it is worth investigating the chain in the absence of solvent so that there is only one excluded volume 
parameter $w_{pp}$ characterizing the monomer-monomer excluded volume interactions. In this particular 
case, the field experienced by monomers is $w_{p}(\mathbf{r}) = w_{pp}\rho_p(\mathbf{r})$. Note that here $w_{pp}$ has the dimensions of volume\cite{edwardsbook} and is 
inversely proportional to the temperature.  
If the local charge per unit volume 
at any instance during the electric charging process is represented by $\rho_{e}'(\mathbf{r}) = \lambda \left [\sum_{j=c,+,-}Z_{j}\rho_{j}'(\mathbf{r})_{\mid_{\lambda \psi'(\mathbf{r})}} + Z_{p}\alpha \rho_{p}'(\mathbf{r})_{\mid_{\lambda \psi'(\mathbf{r}),w_p'(\mathbf{r}) }}\right ]$, where $\psi'(r)$ and 
$w_p'(\mathbf{r}) = w_{pp}'\rho_p'(\mathbf{r})_{\mid_{\lambda \psi'(\mathbf{r}),w_p'(\mathbf{r}) }} $ are the electrostatic potential 
and the field arising from 
excluded volume interactions at the particular instance, then work done ($dF_e$) in 
charging a small volume element $d\Omega = d\mathbf{r}$ by amount $d\lambda$
is given by $dF_{e} = d\lambda d\mathbf{r} \left [\sum_{j=c,+,-}Z_{j}\rho_{j}(\mathbf{r})_{\mid_{\lambda \psi'(\mathbf{r})}} + Z_{p}\alpha \rho_{p}(\mathbf{r})_{\mid_{\lambda \psi'(\mathbf{r}),w_p'(\mathbf{r})}}\right ]k_B T\psi'(\mathbf{r}) = \frac{d\lambda}{\lambda} d\mathbf{r} \rho_{e}'(\mathbf{r})k_B T\psi'(\mathbf{r})$.
So, total work done in charging the whole system from $\lambda = 0 $ to $\lambda = 1 $ is given by
\begin{eqnarray}   
F_{e} &=& \int_{\lambda = 0}^{\lambda = 1} \frac{d\lambda}{\lambda} \int d\mathbf{r}\frac{2 E_{e}'}{\lambda_{e}} \label{eq:fe_charging}
\end{eqnarray}

where 
\begin{eqnarray}
\frac{E_{e}'}{k_B T} &=& \frac{1}{2} \int d \mathbf{r}\rho_{e}'(\mathbf{r})\psi'(\mathbf{r}) 
\end{eqnarray}
is the electrostatic energy at any instance during the electric charging process. 

Similarly, the work done in development of excluded volume parameter $d\lambda^2$ for a 
small volume element $d\Omega$ is given by $dF_{w} = d\mathbf{r} \frac{d\lambda^2}{\lambda^2} \rho_{p}'(\mathbf{r}))_{\mid_{\lambda \psi'(\mathbf{r}), w_p'(\mathbf{r})}}k_B T w_{p}'(\mathbf{r})$,
$k_B$ and $T$ being Boltzmann's constant and temperature, respectively.                              
Hence, the total work done in developing the final value of $w_{pp}$ for the whole system (i.e., from $\lambda = 0 $ 
to $\lambda = 1 $ ) is given by
\begin{eqnarray}
F_{w} &=& \int_{\lambda = 0}^{\lambda = 1} d\lambda^2 \frac{2 E_{w}'}{\lambda^2} \label{eq:fw_charging}
\end{eqnarray}
where 
\begin{eqnarray}
\frac{E_{w}'}{k_B T} &=& \frac{1}{2} \int d\mathbf{r} w_{p}'(\mathbf{r})\rho_{p}'(\mathbf{r})_{\mid_{\lambda \psi'(\mathbf{r}),w_p'(\mathbf{r})}} 
\end{eqnarray}
is the energetic contribution to the free energy at any particular instance arising from the excluded volume interactions. 

From thermodyamics, we know that the free energy ($F$) at any temperature $T$ is related to 
energy ($E$) and entropy ($S$) by the following relations: 

\begin{eqnarray}
F &=& E - T S \\
\Rightarrow \frac{\partial F}{\partial T} &=& -\frac{E}{T^{2}} + \frac{1}{T} \frac{\partial E}{
\partial T} - \frac{\partial S}{\partial T} \nonumber 
\end{eqnarray}

Using $c_{v} = \left ( \frac{\partial E}{\partial T} \right )_{V} = T \left ( \frac{
\partial S}{\partial T} \right )_{V}$
\begin{eqnarray}
\frac{\partial (F/T)}{\partial T}  &=& - \frac{E}{T^{2}} \quad \mbox{and} \quad S =
 - \frac{\partial F}{\partial T} \label{eq:entropy}
\end{eqnarray}

Integrating the last relation between the energy and the free energy, 
\begin{eqnarray}
F &=& T \int_{T = \infty}^{T = T} E  d\left ( \frac{1}{T} \right ) \label{eq:free_thermo}
\end{eqnarray}

It can be shown\cite{marcus55} that the work done as calculated from the \textit{isothermal} charging processes 
 (i.e., $F_e + F_w$) is related to the free energy ($F$) calculated by integrating energy $E$ over the
inverse temperature (cf. Eq. ~\ref{eq:free_thermo}) and in general, $F_e + F_w $ is not equal to $F$.
For the two processes considered here, the total work done during isothermal charging ($F_{ew}$) is 
given by $F_{ew} = F_{e} + F_{w}$. $F_{ew}$ can be related to $F$ by noting that the expression for the 
energy at any instance is the same in the two descriptions i.e., $E' = E_e' + E_w'$, because from Eq. (~\ref{eq:free_thermo}), 

\begin{eqnarray}
F' &=& T \int_{T = \infty}^{T = T} E'  d\left ( \frac{1}{T} \right ) \label{eq:free_thermo_instant}\\
\Rightarrow \left(\frac{\partial (F'/T)}{\partial 1/T}\right)_\lambda  &=& E' = E_e' + E_w'. \label{eq:deri_thermo_t}
\end{eqnarray}

Also, from Eqs. (~\ref{eq:fe_charging}) and (~\ref{eq:fw_charging}), 
\begin{eqnarray}
\left(\frac{\partial F_e'/T}{\partial \lambda}\right)_T &=& \frac{2E_e'}{\lambda T} \label{eq:deri_charging_ee}
\end{eqnarray}
and 
\begin{eqnarray}
\left(\frac{\partial F_w'/T}{\partial \lambda^2}\right)_T &=& \frac{2E_w'}{\lambda^2 T}. \label{eq:deri_charging_ww}
\end{eqnarray}
Here, $F_e'$ and $F_w'$ are the work done during the electric and excluded volume 
charging at any instant characterized by a particular value of the charging parameter $\lambda$. 

From Eqs. (~\ref{eq:deri_thermo_t}), (~\ref{eq:deri_charging_ee}) and (~\ref{eq:deri_charging_ww}), 

\begin{eqnarray}
\frac{2}{\lambda T}\left(\frac{\partial (F'/T)}{\partial \left(1/T\right)}\right)_\lambda &=& \left(\frac{\partial (F_e' + F_w'/2)/T}{\partial \lambda}\right)_T.  \label{eq:relate_methods}
\end{eqnarray}
It can be shown that Eq. (~\ref{eq:relate_methods}) is satisfied as long as $(F_e' + F_w'/2)/T$ 
is related to $\lambda$ and $T$ only through $\lambda^{2}/T$ at fixed dielectric
constant $\epsilon$ (after noting that $\frac{F}{T} =  \int_{-\infty}^{\ln(1/T)} d \left (\ln \left [ \frac{\lambda^{2}}{T}\right ] \right ) \frac{E'}{T} $). Marcus has already shown that the
use of the Poisson-Boltzmann equation for computing the electrostatic potential does not 
lead to the violation of the constraint presented in Eq. (~\ref{eq:relate_methods}). 
Also, this particular constraint for relating the work done from 
the isothermal charging process to the free energy has led us to choose $\lambda^2$ as the prefactor 
for instant value of the excluded volume parameter so that the instant values of the monomer 
densities and fields are dependent on $\lambda^2/T$ (after taking the 
excluded volume parameter $w_{pp}$ to be of the form $a/T$, $a$ being a constant independent of temperature $T$). 
Furthermore, note that if the instant value of excluded volume parameter is taken to 
be $w_{pp}' = \lambda w_{pp}$ similar 
to the electric charging process, the computation of the monomer density by solving the modified diffusion equation would
lead to the violation of the constraint given by Eq. (~\ref{eq:relate_methods}) and make it difficult to deduce the 
energetic and entropic contributions to the free energy. Also, Eq. (~\ref{eq:relate_methods}) is part of the reason that the dielectric constant of the medium has to be assumed to be independent of temperature, while carrying out 
this analysis.  

Having shown that the work done during the isothermal charging process can be used to compute the free energy of the 
system, it is clear that for the dual charging process considered here, they are related by $F = F_e + F_w/2$. 
Using this relation between the work done as a result of the charging and the free energy, the entropy can 
be calculated using Eq. (~\ref{eq:entropy}) so that 

\begin{eqnarray}
-S &=& \frac{\partial F}{\partial T} = \int_{0}^{1} \frac{d \lambda}{\lambda} \frac{\partial \left[ 2 E'\right ]}{\partial T} \\
&=& \int_{0}^{1} \frac{d \lambda}{\lambda} \frac{\partial \left[ 2 E_{e}'\right ]}{\partial T} + \int_{0}^{1} \frac{d \lambda^2}{\lambda^2} \frac{\partial \left[E_{w}'\right ]}{\partial T} \\
&=&  - S_{e} - S_{w} 
\end{eqnarray}
where
\begin{eqnarray}
- S_{e} &=& \int_{0}^{1} \frac{d \lambda}{\lambda} \frac{\partial}{\partial T}\left \{ \int d\mathbf{r} \rho_e'(\mathbf{r})k_B T\psi'(\mathbf{r})\right \} \label{eq:entropy_elec}\\
- S_{w} &=& \int_{0}^{1} \frac{d \lambda}{\lambda} \frac{\partial }{\partial T}\left \{ \int d\mathbf{r} k_B T w_{p}'(\mathbf{r})\rho_{p}'(\mathbf{r})\right \} \label{eq:entropy_excl}
\end{eqnarray}
Now, using the integral form for the Poisson equation obtained at the saddle point $\psi'(\mathbf{r}) = - 4\pi l_B \int d\mathbf{r}' \rho_e'(\mathbf{r}')/|\mathbf{r}'|$, the assumption that $\epsilon$ is independent of 
$T$ and $w_{pp}' = a/T$, the above equations take the form 
\begin{eqnarray}
- S_{e} &=& 2\int d\mathbf{r}\int_{0}^{1} \frac{d \lambda}{\lambda} k_B T\psi'(\mathbf{r})\frac{\partial \rho_e'(\mathbf{r})}{\partial T} \label{eq:entropy_elec_2}\\
- S_{w} &=& 2\int d\mathbf{r} \int_{0}^{1} \frac{d \lambda}{\lambda} k_B T w_{p}'(\mathbf{r})\frac{\partial \rho_{p}'(\mathbf{r})}{\partial T} \label{eq:entropy_excl_2}
\end{eqnarray}

Now, using an identity for any arbitrary function $f$ of $\lambda^{2}/T$
\begin{eqnarray}                                                                
\lambda \frac{\partial }{\partial T}\left[ f\left(\frac{\lambda^{2}}{T}\right)\right ] + \frac{\lambda^{2}}{2T} \frac{\partial }{\partial \lambda}\left[ f\left(\frac{\lambda^{2}}{T}\right)\right ] &=& 0 \label{eq:identity_lam}
\end{eqnarray}  
expressions for the entropic contributions can be cast in the form
\begin{eqnarray}
- S_{e} &=& -k_B\int d\mathbf{r}\int_{0}^{1} d \lambda \lambda \psi'(\mathbf{r})\frac{\partial }{\partial \lambda}\left\{\frac{\rho_e'(\mathbf{r})}{\lambda}\right\} \label{eq:entropy_elec_3}\\
- S_{w} &=& - k_B\int d\mathbf{r} \int_{0}^{1} d \lambda  w_{p}'(\mathbf{r})\frac{\partial \rho_{p}'(\mathbf{r})}{\partial \lambda} \label{eq:entropy_excl_3}
\end{eqnarray}

Carrying out integration by parts over $\lambda$
\begin{eqnarray}
- S_{e} &=& - k_B \int d\mathbf{r} \psi(\mathbf{r})\rho_e(\mathbf{r})
+ k_B \int d\mathbf{r} \int_{0}^{1} d \lambda \frac{\rho_e'(\mathbf{r})}{\lambda}\frac{\partial \left\{\lambda \psi'(\mathbf{r})\right \}}{\partial \lambda} \label{eq:entropy_elec_4}\\
- S_{w} &=& - k_B\int d\mathbf{r} w_{p}(\mathbf{r})\rho_{p}(\mathbf{r}) 
+ k_B\int d\mathbf{r} \int_{0}^{1} d \lambda  \rho_{p}'(\mathbf{r})\frac{\partial w_{p}'(\mathbf{r})}{\partial \lambda} \label{eq:entropy_excl_4}
\end{eqnarray}
Here, we have used the fact that the fields and charge density are zero for $\lambda = 0$ i.e., $\psi'(\mathbf{r}) = \rho_e'(\mathbf{r}) = w_p'(\mathbf{r}) = 0$ for $\lambda = 0$. Realizing that the partition functon for a 
single small ion can be used to rewrite the entropy by 

\begin{eqnarray}
\frac{\partial}{\partial \lambda} \left\{n_{j}\ln Q_j'\right\} &=& \frac{\partial}{\partial \lambda} \left\{n_{j}\ln \left [\int d\mathbf{r} \exp \left( - Z_{j} \lambda \psi'(\mathbf{r})\right ) \right ] \right \} \nonumber \\
&=& - \int d\mathbf{r} Z_j \rho_{j}'(\mathbf{r})\frac{\partial \left\{\lambda \psi'(\mathbf{r})\right \}}{\partial \lambda} \label{eq:logqj}
\end{eqnarray}
where 
\begin{eqnarray}
\rho_{j}'(\mathbf{r}) &=& \frac{n_{j}}{Q_j}\exp \left( - Z_{j} \lambda \psi'(\mathbf{r})\right ) \label{eq:ion_density_instant}\\
Q_j &=& \int d\mathbf{r}\exp \left( - Z_{j} \lambda \psi'(\mathbf{r})\right ). 
\end{eqnarray}

Similarly, the entropic contributions from the polymer can be rewritten using 

\begin{equation}
Q_{p}' = \int d\mathbf{r} \int d\mathbf{r}'' G'(\mathbf{r},0,\mathbf{r}'',N)
\end{equation}
where
\begin{eqnarray}
G'(\mathbf{r},0,\mathbf{r}'',N)&=& \int_{\mathbf{r}}^{\mathbf{r}''} D[\mathbf{R}]\exp \left [-\frac {3 }{2 b^{2}} \int_{0}^{N} dt \left(\frac{\partial \mathbf{R}(t)}{\partial t} \right )^{2} - \int_{0}^{N} {dt} \left \{ Z_{p}\alpha \lambda \psi'(\mathbf{R}(t)) + w_{p}'(\mathbf{R}(t))\right \} \right ]\nonumber \\
 && \delta\left [\mathbf{r}-\mathbf{R}(0)\right ] \delta\left [\mathbf{r}''-\mathbf{R}(N)\right ] 
\end{eqnarray}
Now, 
\begin{eqnarray}
\frac{\partial \ln Q_{p}'}{\partial \lambda} &=& \frac{1}{Q_{p}'}\frac{\partial Q_{p}'}{\partial \lambda}  = \frac{1}{Q_p'}\int d\mathbf{r} \int d\mathbf{r}''\frac{\partial G'(\mathbf{r},0,\mathbf{r}'',N)}{\partial \lambda} \\
&=&  - \int d\mathbf{r}'\rho_{p}'(\mathbf{r}')\mid_{\lambda \psi'(\mathbf{r}'), w_{p}'(\mathbf{r}')} \frac{\partial}{\partial \lambda} \left \{ \lambda Z_{p}\alpha \psi'(\mathbf{r}') + w_{p}'(\mathbf{r}') \right \} \label{eq:logqp}
\end{eqnarray}

Using Eqs. (~\ref{eq:logqj}) and (~\ref{eq:logqp})
\begin{eqnarray}
-S = - S_{e} - S_w &=& - k_B \int d\mathbf{r} \psi(\mathbf{r})\rho_e(\mathbf{r}) - k_B\int d\mathbf{r} w_{p}(\mathbf{r})\rho_{p}(\mathbf{r})\nonumber \\
&& - k_B \sum_{j=c,+,-} n_j \int_{0}^{1} d \lambda  \frac{\partial \ln Q_j'}{\partial \lambda}  
- k_B \int_{0}^{1} d \lambda  \frac{\partial \ln Q_p'}{\partial \lambda}. \label{eq:entropy_total_2}
\end{eqnarray}
Carrying out the integrals over $\lambda$, we get

\begin{eqnarray}
-S &=&  - k_B \int d\mathbf{r} \psi(\mathbf{r})\rho_e(\mathbf{r}) - k_B\int d\mathbf{r} w_{p}(\mathbf{r})\rho_{p}(\mathbf{r})\nonumber \\
&& - k_B \sum_{j=c,+,-} n_j\ln \left[\frac{Q_j}{\Omega}\right]  
- k_B \ln \left[\frac{Q_p}{Q_p'\left\{\lambda = 0\right\}}\right], \label{eq:entropy_total_3}
\end{eqnarray}
where we have used the fact that $Q_j' = \Omega$ for $\lambda =  0$. An important point to note here is that 
$S$ turns out to be the entropy of the system \textit{relative} to the ideal system (i.e., a system 
in the absence of interactions). In other words, $F = E - T S$ where $S = S_{total}-S_{id}$, as was 
already pointed out by Marcus in Ref. \cite{marcus55}.

Now, it is easy to identify the entropic contributions arising from different components. 
Separating $S$ into contributions from small ions and the polymer chain by writing
\begin{eqnarray}
- S &=&  - \left(S_{ions}-S_{ions}^{id}\right) - \left(S_{poly}-S_{poly}^{id} \right)\\
- \left(S_{ions}-S_{ions}^{id}\right) &=& - k_B \sum_{j=c,+,-} \left [ \int d\mathbf{r} Z_j\rho_j(\mathbf{r})\psi(\mathbf{r}) + n_j\ln \left[\frac{Q_j}{\Omega}\right]  \right ]\\
&=& k_B\sum_{j=c,+,-}  \left[ \int d\mathbf{r} \rho_j(\mathbf{r})\left\{\ln \rho_j(\mathbf{r}) - 1\right\} - n_j \left\{\ln \frac{n_j}{\Omega} - 1\right \} \right] \label{eq:entropy_ions_final}\\
-\left(S_{poly} -S_{poly}^{id}\right)&=&  - k_B \int d\mathbf{r} Z_p \alpha \rho_p(\mathbf{r}) \psi(\mathbf{r}) - k_B\int d\mathbf{r} w_{p}(\mathbf{r})\rho_{p}(\mathbf{r})- k_B \ln \left[\frac{Q_p}{Q_p'\left\{\lambda = 0\right\}}\right ]
\nonumber \\
&& \label{eq:entropy_poly_final}
\end{eqnarray}
where quantities with ``$id$'' in the superscripts are the entropic contributions in the ideal system. 
In writing the entropy of small ions in terms of densities, we have used Eq. (~\ref{eq:ion_density_instant}) after 
putting $\lambda = 1$. Also, note that summing up the energy and entropy, the total free energy 
of the system obtained using the charging method described here differs from the free energy 
obtained within the saddle point approximation of SCFT by the entropic contributions 
of the ideal system in the absence of interactions, which is the most obvious choice for the 
reference frame during the computations of free energy in the field theory. In particular, the free 
energy of the reference state comes out to be 
\begin{eqnarray}
\frac{F_{ref}}{k_{B}T} &=& -S_{ions}^{id} - S_{poly}^{id} \nonumber \\
& = & \sum_{j=c,+,-} n_{j}\left [ \ln \frac{n_{j}}{\Omega} - 1\right ] - \ln Q_{p}'\left\{\lambda = 0\right\}
\end{eqnarray}

\renewcommand{\theequation}{C-\arabic{equation}}
  \setcounter{equation}{0}  
  \section*{APPENDIX C : Energy and entropy of single polyelectrolyte chain - incompressible system}
 
In Appendix B, we presented the derivation for the energetic and entropic contributions to the single 
polyelectrolyte chain system in the absence of solvent. Here, we present the  
generarlization of the technique to polyelectrolyte chain in the presence 
of solvent treated within the incompressibility constraint and point-like ions 
(i.e., $\rho_p(\mathbf{r}) + \rho_s(\mathbf{r}) = \rho_0$). 
In this case, there are three excluded volume parameters $w_{pp}, w_{ss}$ and $w_{ps}$ characterizing 
monomer-monomer, solvent-solvent and monomer-solvent interactions in contrast to just one in Appendix B. 
Expressions for the electrostatic contributions remain the same and here, we focus on the 
contributions arising from the excluded volume interactions. 

Consider the two isothermal charging processes imagined in Appendix B i.e., 
the traditional electric and excluded volume ``charging'', where  charge of the ionic species 
and the excluded volume parameters characterizing the 
short range excluded volume effects are developed linearly and quadratically from $0$ to their final values 
in term of the charging parameter $\lambda$. At any instance the excluded volume 
parameter between species $i$ and $j$ is $w_{ij}' = \lambda^2 w_{ij}$, 
where $w_{ij}$ is the final value of the excluded volume parameter. Also, the incompressibility 
constraint is forced at all instances during the charging process. 

Excluded volume energy for the system can be written as 
    
\begin{eqnarray}
\frac{E_{w}}{k_B T} &=& \frac{1}{2} \int d\mathbf{r} \left[w_{pp}\rho_p^2(\mathbf{r}) + w_{ss}\rho_{s}^2(\mathbf{r}) + 2w_{ps}\rho_p(\mathbf{r}) \rho_s(\mathbf{r})\right]\\
&=& \frac{\rho_0}{2}\left[w_{pp}N + w_{ss}n_s\right] + \chi_{ps}b^3\int d\mathbf{r} \rho_p(\mathbf{r})\rho_s(\mathbf{r})
\end{eqnarray}
where $\chi_{ps}b^3 = w_{ps} - (w_{pp} + w_{ss})/2$ is the Flory's chi parameter and it is taken to be of the form $a/T$ from the 
assumed dependence of the excluded volume parameters $w_{ij}$. 

Following the recipe presented in Appendix B, the effect of excluded volume interactions on the entropic contributions 
can be written as  

\begin{eqnarray}
- S_{w} &=& 2\int_{0}^{1} \frac{d \lambda}{\lambda} \frac{\partial }{\partial T}\left \{ \int d\mathbf{r} k_B T 
\chi_{ps}'b^3\rho_p'(\mathbf{r})\rho_{s}'(\mathbf{r})\right \} \label{eq:entropy_excl_incompress}\\
&=& 2k_B T \chi_{ps}'b^3 \int d\mathbf{r} \int_{0}^{1} \frac{d \lambda}{\lambda} \left\{\rho_{p}'(\mathbf{r})\frac{\partial \rho_{s}'(\mathbf{r})}{\partial T}  + \rho_{s}'(\mathbf{r})\frac{\partial \rho_{p}'(\mathbf{r})}{\partial T} \right \}\label{eq:entropy_excl_2_incompress}\\
&=& 2k_B T \int d\mathbf{r} \int_{0}^{1} \frac{d \lambda}{\lambda} \left\{w_{s}'(\mathbf{r})\frac{\partial \rho_{s}'(\mathbf{r})}{\partial T}  + w_{p}'(\mathbf{r})\frac{\partial \rho_{p}'(\mathbf{r})}{\partial T} \right \},
\end{eqnarray}
where we have used $w_p'(\mathbf{r}) = \chi_{ps}'b^3 \rho_s'(\mathbf{r}) + \eta'(\mathbf{r}),w_s'(\mathbf{r}) = \chi_{ps}'b^3 \rho_p'(\mathbf{r}) + \eta'(\mathbf{r}) $ after adding $\eta'(\mathbf{r})\frac{\partial \left\{\rho_P'(\mathbf{r}) + \rho_s'(\mathbf{r})\right\}}{\partial T} = 0$ in the last step. 
Using the procedure described in Appendix B
\begin{eqnarray}
-S &=&  - k_B \int d\mathbf{r} \psi(\mathbf{r})\rho_e(\mathbf{r}) - k_B\int d\mathbf{r} w_{p}(\mathbf{r})\rho_{p}(\mathbf{r}) - k_B\int d\mathbf{r} w_{s}(\mathbf{r})\rho_{s}(\mathbf{r})\nonumber \\
&& - k_B \sum_{j=c,+,-} n_j\ln \left[\frac{Q_j}{\Omega}\right] - k_B n_s\ln \left[\frac{Q_s}{\Omega}\right] 
- k_B \ln \left[\frac{Q_p}{Q_p'\left\{\lambda = 0\right\}}\right], \label{eq:entropy_total_3_incompress}
\end{eqnarray}
where we have defined $Q_s' = \int d\mathbf{r}\exp\left[-w_s'(\mathbf{r})\right]$ and $\rho_s'(\mathbf{r}) = n_s \exp\left[-w_s'(\mathbf{r})\right]/Q_s'$. An extra contribution due to the solvent appears in the expression 
for entropy so that 
\begin{eqnarray}
  - S &=&  - \left(S_{ions}-S_{ions}^{id}\right) -\left(S_{solvent}-S_{solvent}^{id}\right) - \left(S_{poly}-S_{poly}^{id} \right)\\
-\left(S_{solvent}-S_{solvent}^{id}\right)  &=& - k_B \left[\int d\mathbf{r} \rho_s(\mathbf{r})w_s(\mathbf{r}) + n_s\ln \left\{\frac{Q_s}{\Omega}\right\}\right] \\
&=& k_B \left[\int d\mathbf{r} \rho_s(\mathbf{r})\left\{\ln \rho_s(\mathbf{r}) - 1\right\} - n_s \left\{\ln \frac{n_s}{\Omega} - 1\right \} \right] 
\end{eqnarray}
and the expressions for $S_{ions}-S_{ions}^{id}$ and $S_{poly}-S_{poly}^{id}$ are the same as in Eqs. (~\ref{eq:entropy_ions_final}) and 
(~\ref{eq:entropy_poly_final}), respectively.

\newpage
\section*{FIGURE CAPTION}
\pagestyle{empty}
\begin{description}
\item[Fig. 1.:] Cartoons of the chain in the ``free ends'' (1a) and ``end-fixed'' (1b) states. Spherical coordinates used to solve the SCFT equations are shown in Fig. 1c.
\end{description}

\begin{description}
\item[Fig. 2.:] (Color) Two dimensional monomer and electrostatic potential distribution for the single flexible polyelectrolyte chain in ``free ends'' $[(a)$ and $(b)$, respectively $]$ and ``end-fixed'' state $[(c)$ and $(d)$, respectively$]$. $l_{B}/b = 3, \alpha = 0.1, c_{s} = 0.1M, N = 50, R/b = 4$ and $\chi_{ps} = 0.45$. For plots $(c)$ and $(d)$, one end is anchored at $\left[x,y\right] = \left[(R-0.625)b,0\right]$, which is shown by an arrow.
\end{description}

\begin{figure}[ht!]
\vspace*{-0.95cm}
\begin{center}
(a)$\begin{array}{c@{\hspace{.001in}}c@{\hspace{.001in}}}
\includegraphics[width=3in,height=3in]{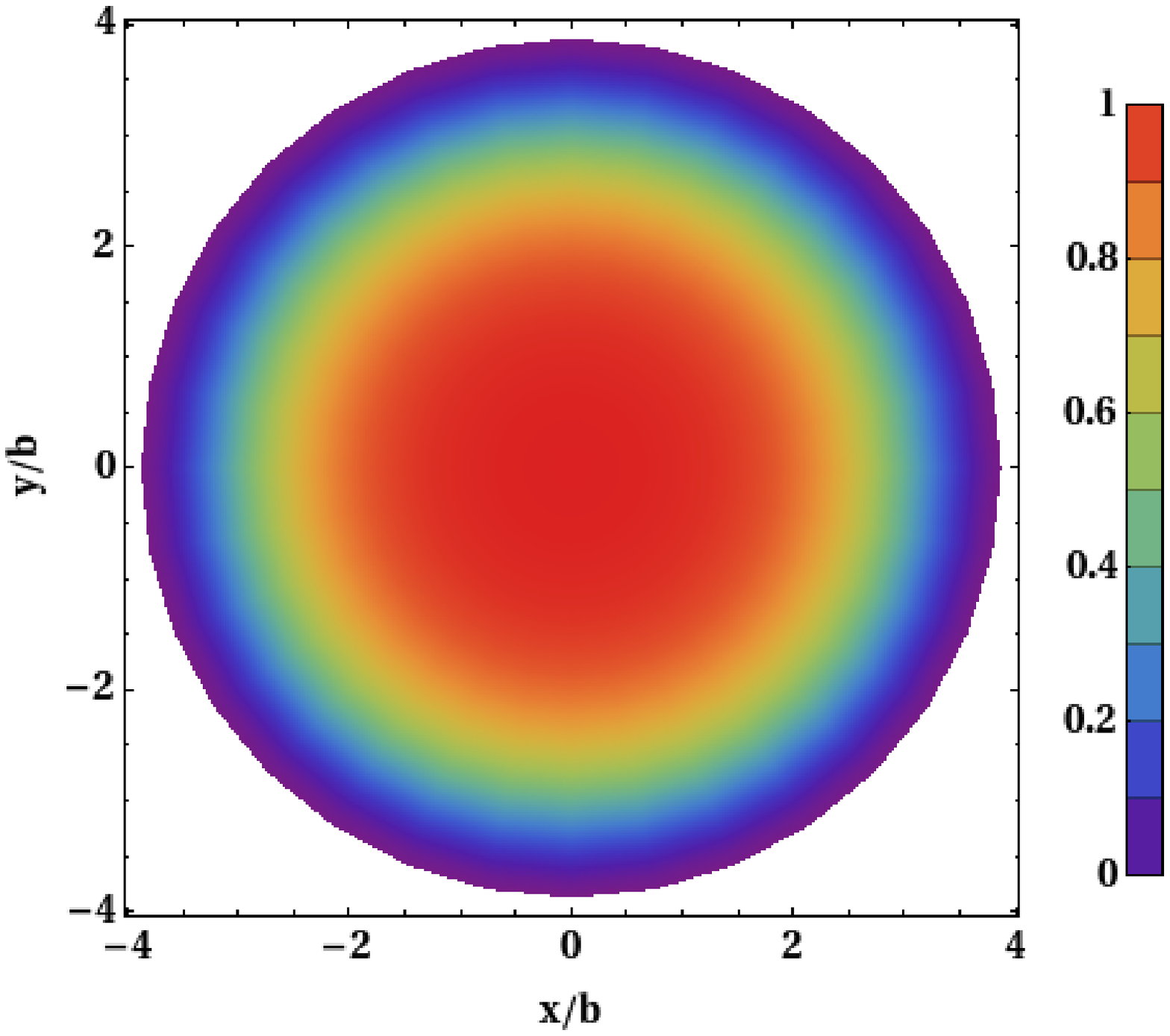}  &
\includegraphics[width=3in,height=3in]{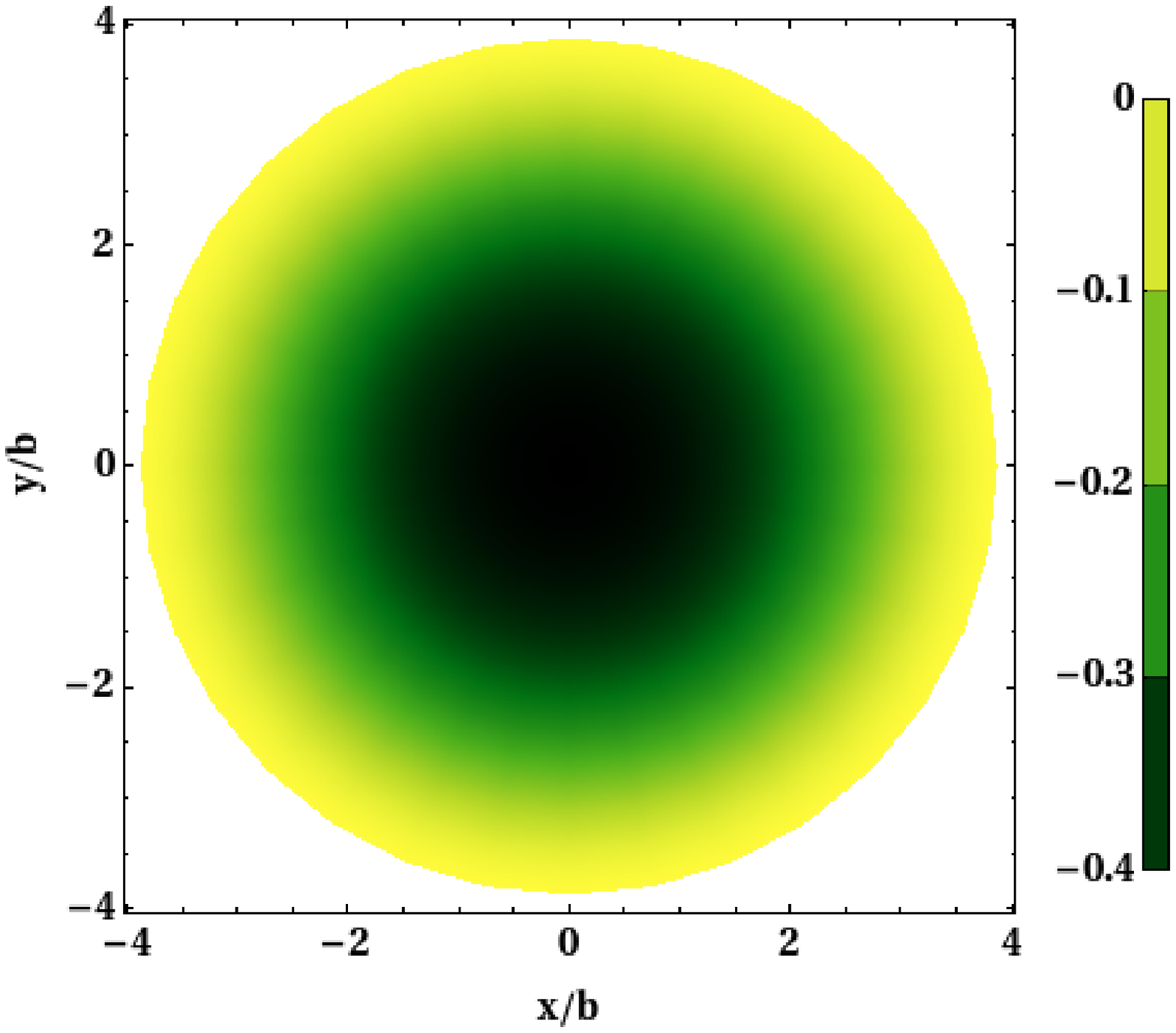}
\end{array}$(b)
\newline
(c)$\begin{array}{c@{\hspace{.001in}}c@{\hspace{.001in}}}
 \includegraphics[width=3in,height=3in]{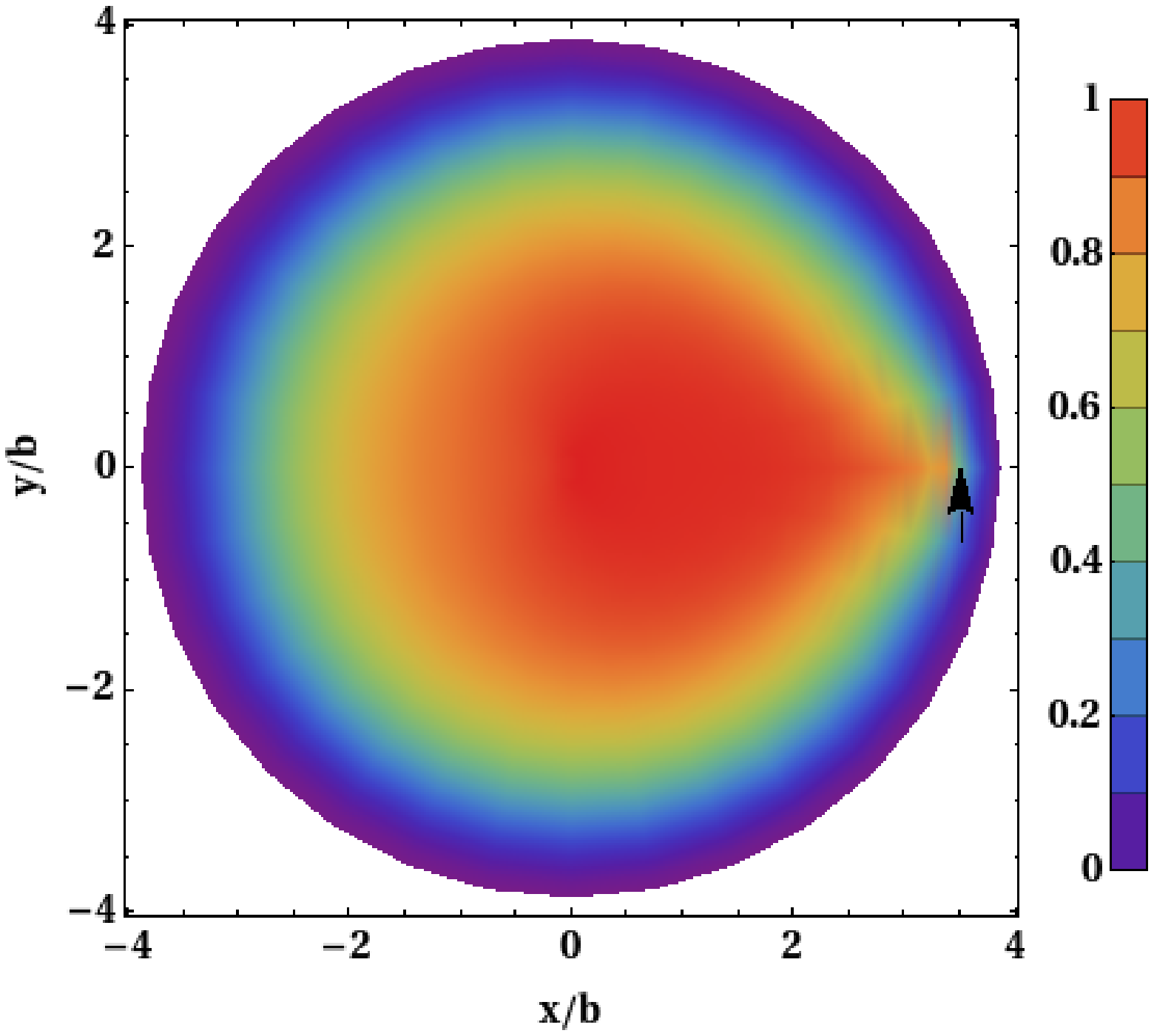} &
 \includegraphics[width=3in,height=3in]{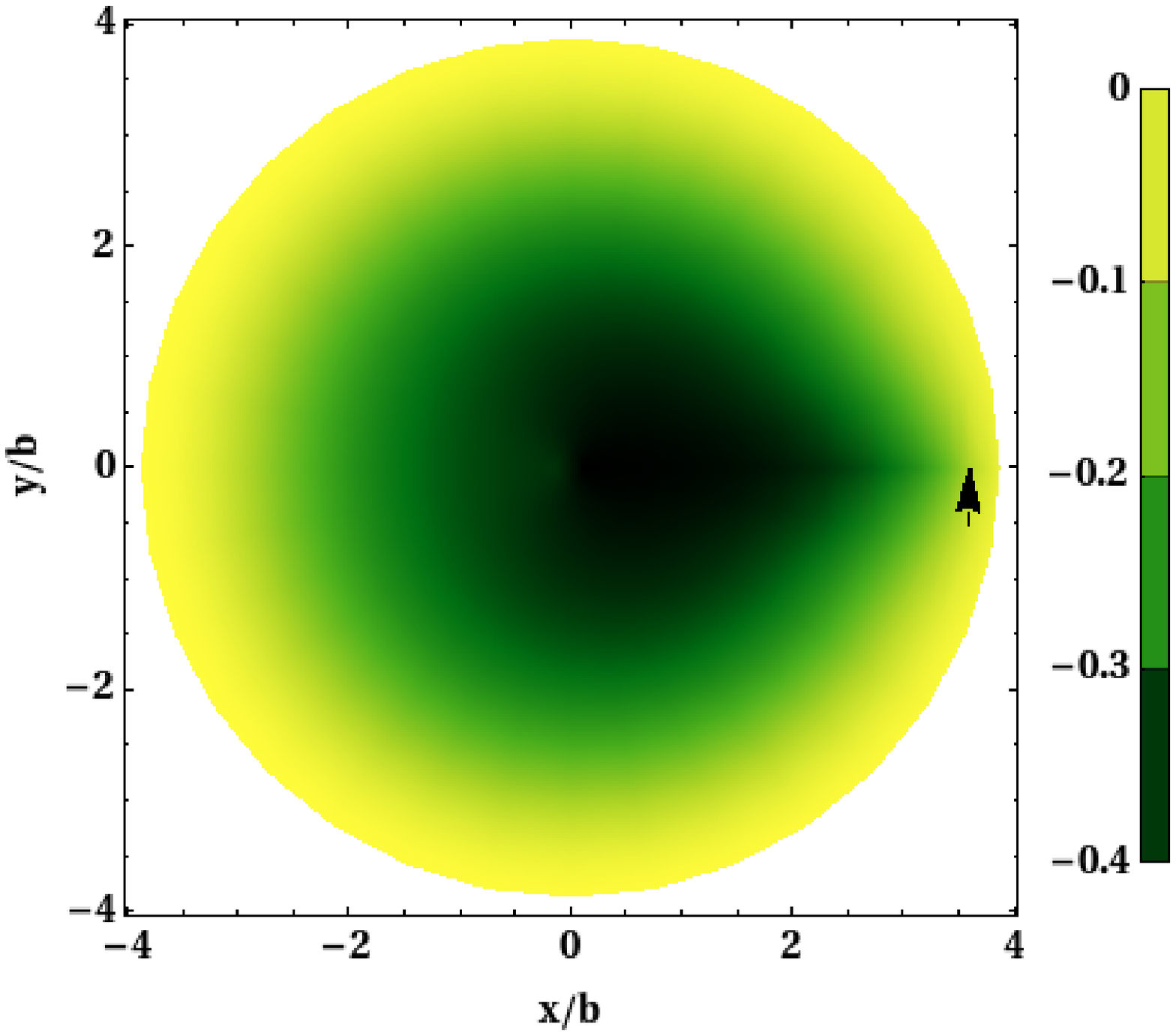} 
\end{array}$(d)
\vspace*{-1.0cm}
\newline
\end{center}
\caption{(Color) Two dimensional monomer and electrostatic potential distribution for the single flexible polyelectrolyte chain in ``free ends'' $[(a)$ and $(b)$, respectively $]$ and ``end-fixed'' state $[(c)$ and $(d)$, respectively$]$. $l_{B}/b = 3, \alpha = 0.1, c_{s} = 0.1M, N = 50, R/b = 4$ and $\chi_{ps} = 0.45$. For plots $(c)$ and $(d)$, one end is anchored at $\left[x,y\right] = \left[(R-0.625)b,0\right]$, which is shown by an arrow.}
\label{fig:den_pot_n50}
\end{figure}

\begin{figure}[ht!]
\vspace*{-0.95cm}
\begin{center}
(a)$\begin{array}{c@{\hspace{.001in}}c@{\hspace{.001in}}}
\includegraphics[width=3in,height=3in]{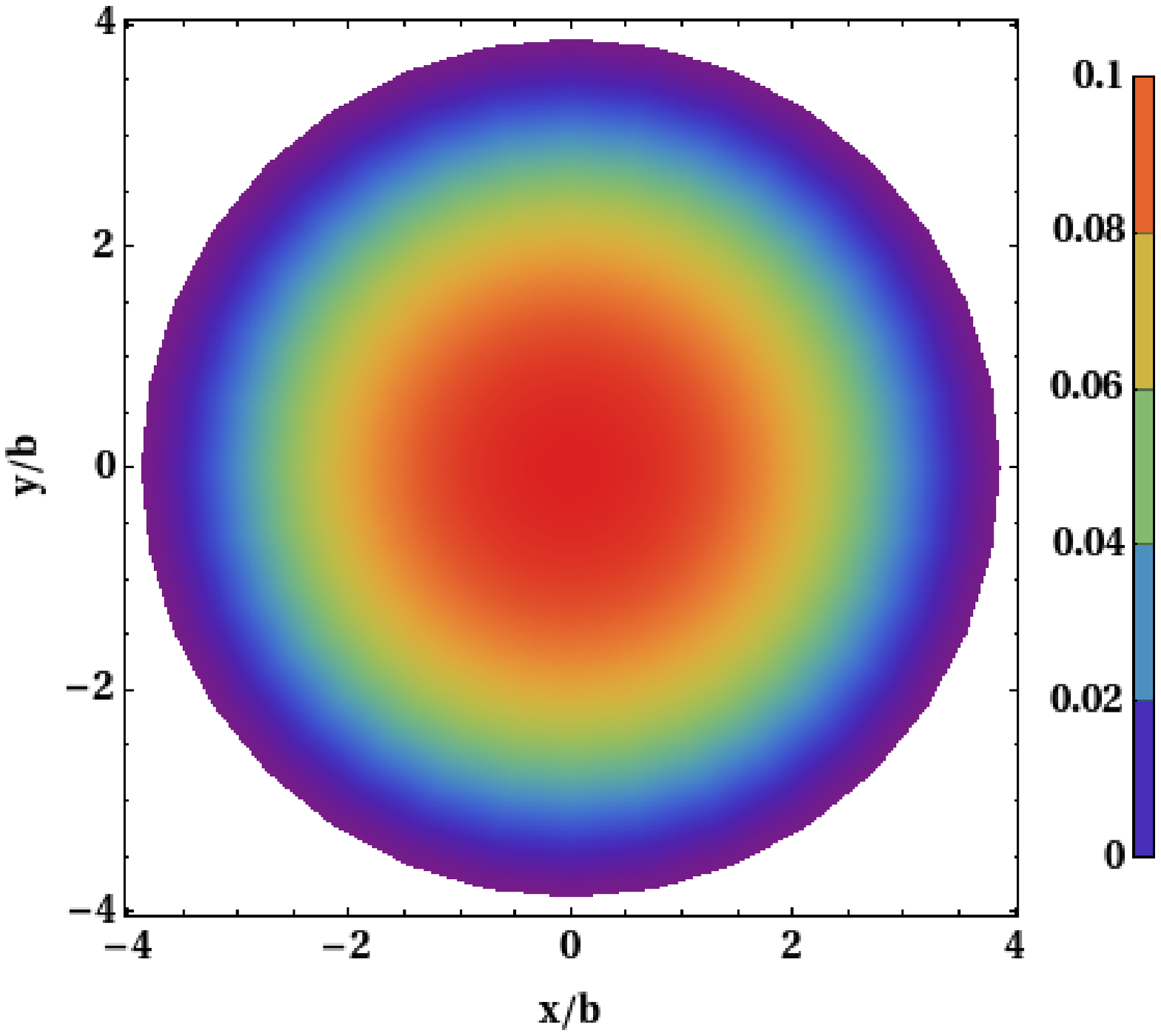}  &
\includegraphics[width=3in,height=3in]{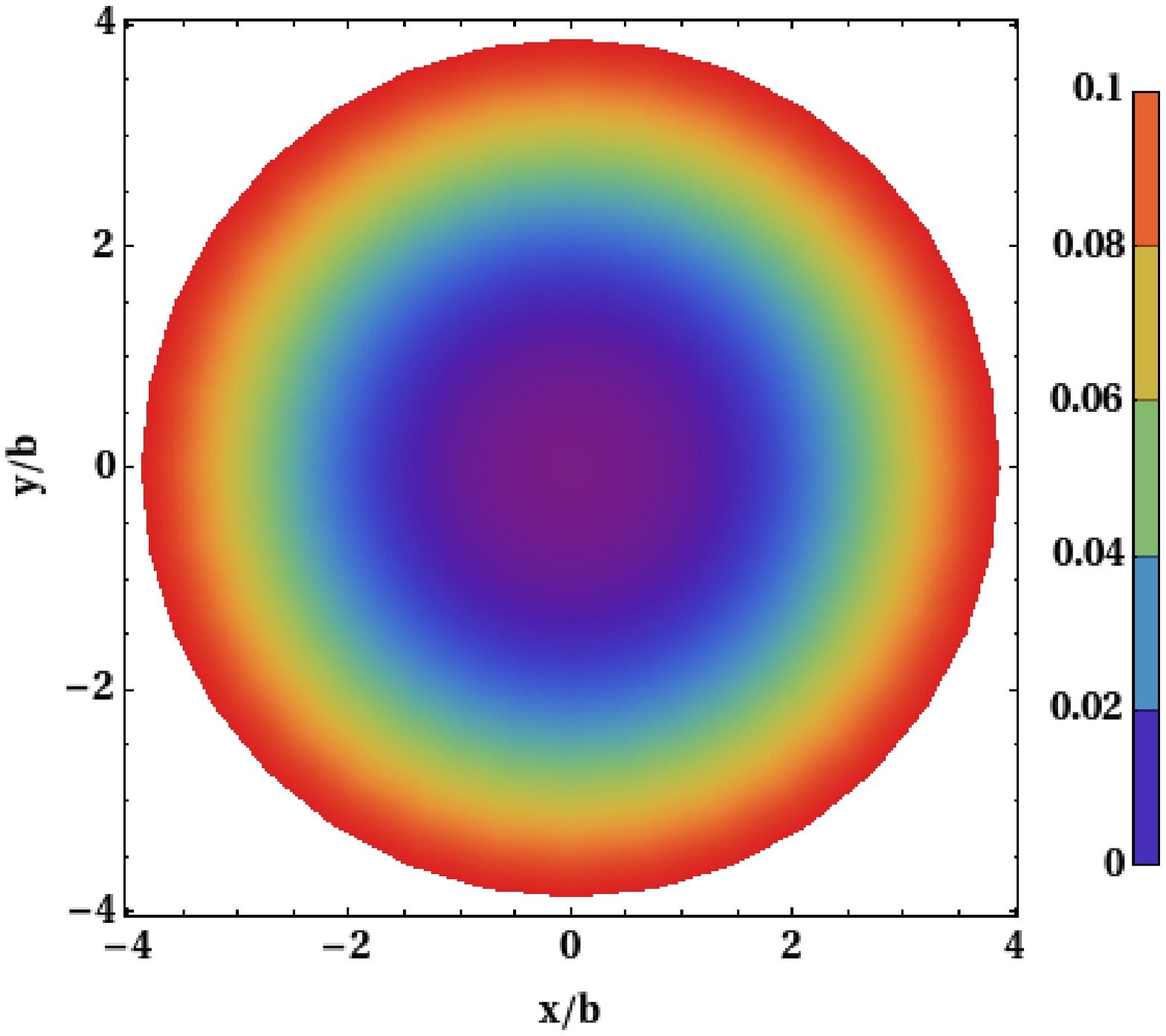}
\end{array}$(b)
\newline
(c)$\begin{array}{c@{\hspace{.001in}}c@{\hspace{.001in}}}
 \includegraphics[width=3in,height=3in]{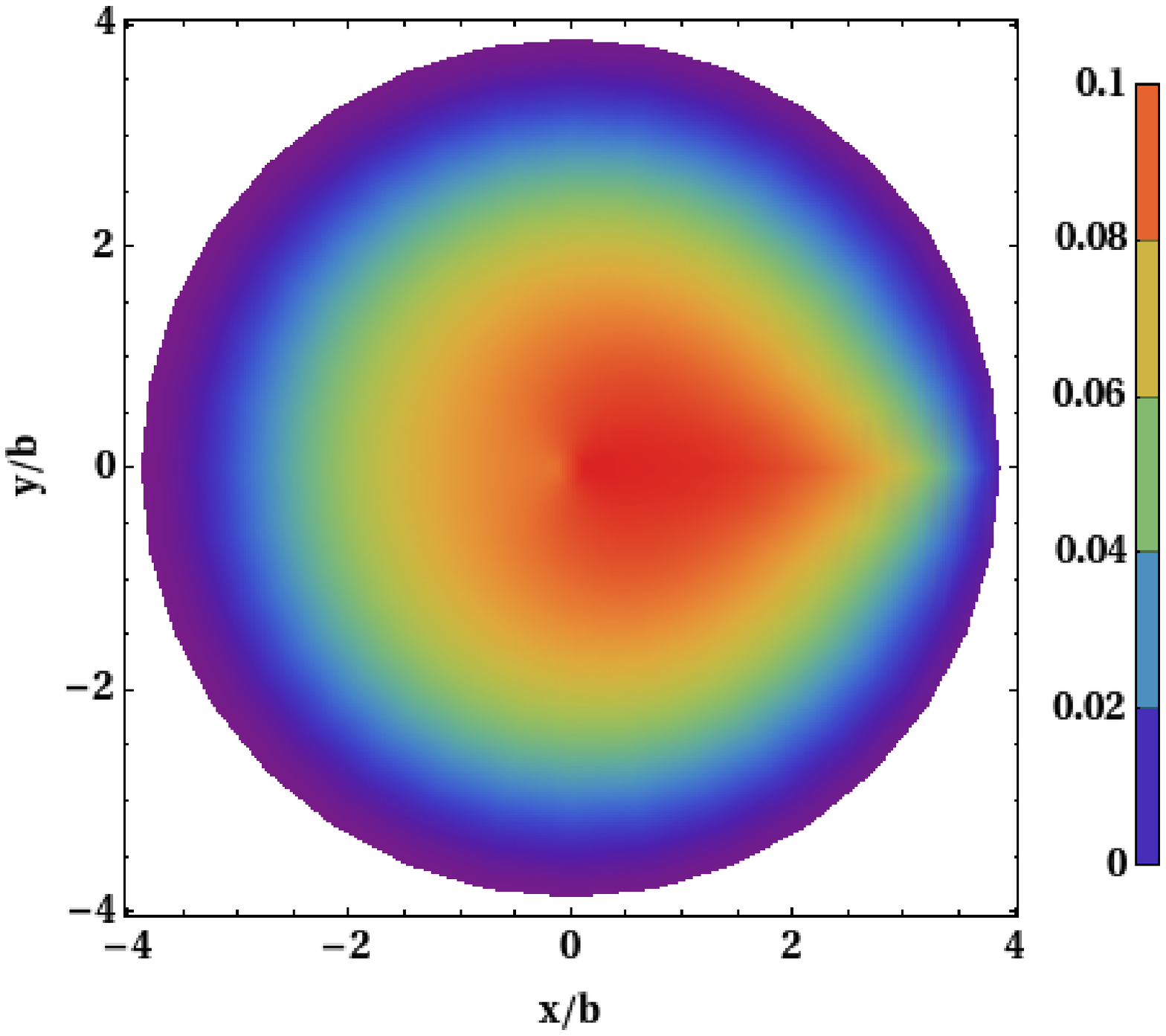} &
 \includegraphics[width=3in,height=3in]{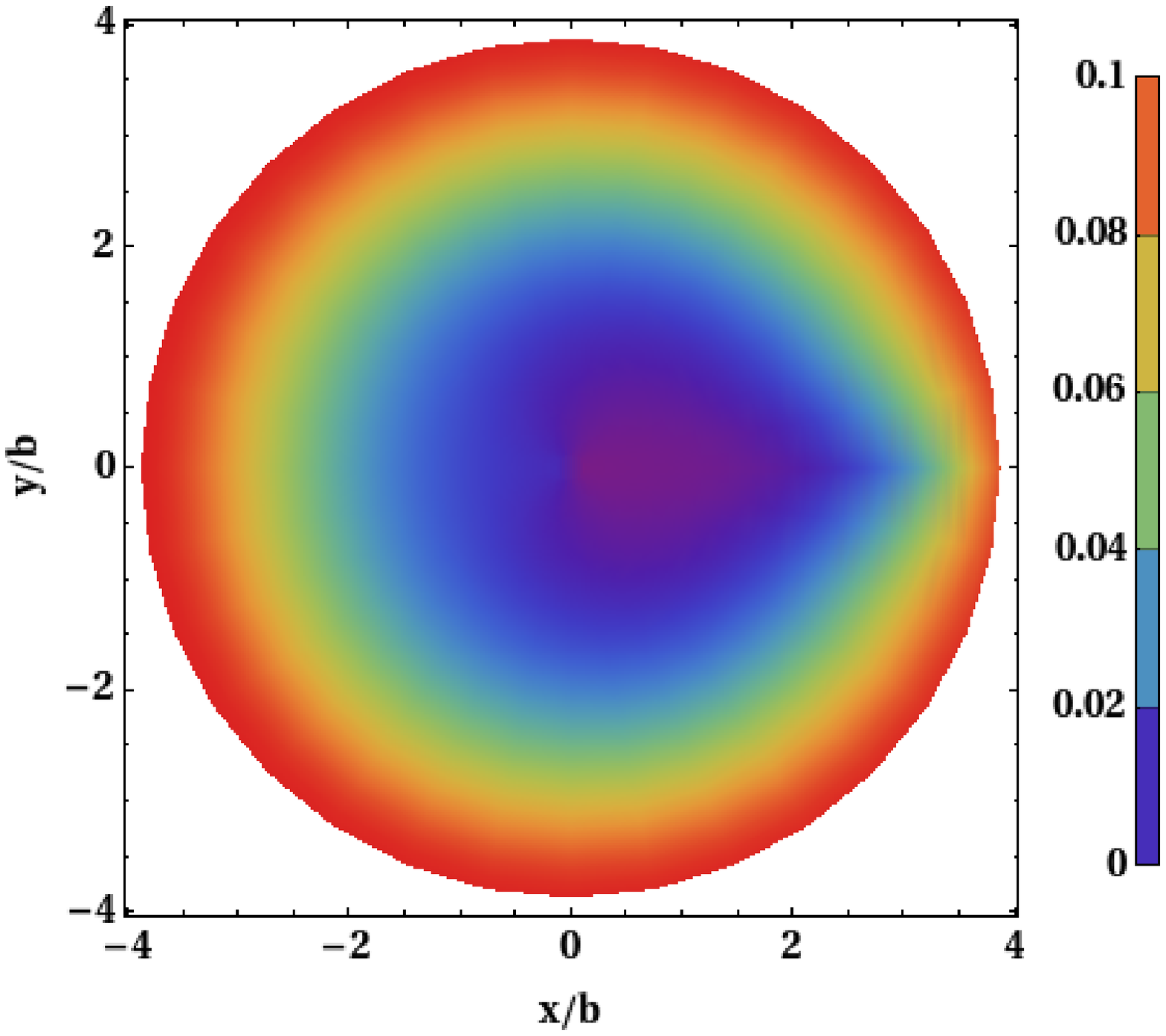}
\end{array}$(d)
\vspace*{-1.0cm}
\newline
\end{center}
\caption{(Color) Counterion and coion distribution for the single flexible polyelectrolyte chain in ``free ends'' $[(a)$ and $(b)$, respectively $]$ and ``end-fixed'' state $[(c)$ and $(d)$, respectively$]$. All the parameters are the same as in Fig. ~\ref{fig:den_pot_n50} i.e., $l_{B}/b = 3, \alpha = 0.1, c_{s} = 0.1M, N = 50, R/b = 4$ and $\chi_{ps} = 0.45$. For the ``end-fixed'' state in plots $(c)$ and $(d)$, one end is anchored at $\left[x,y\right] = \left[(R-0.625)b,0\right]$.} \label{fig:ions_n50}
\end{figure}

\begin{description}
\item[Fig. 3.:](Color) Counterion and coion distribution for the single flexible polyelectrolyte chain in ``free ends'' $[(a)$ and $(b)$, respectively $]$ and ``end-fixed'' state $[(c)$ and $(d)$, respectively$]$. All the parameters are the same as in Fig. ~\ref{fig:den_pot_n50} i.e., $l_{B}/b = 3, \alpha = 0.1, c_{s} = 0.1M, N = 50, R/b = 4$ and $\chi_{ps} = 0.45$. For the ``end-fixed'' state in plots $(c)$ and $(d)$, one end is anchored at $\left[x,y\right] = \left[(R-0.625)b,0\right]$.
\end{description}

\begin{description}
\item[Fig. 4.:] (Color) Monomer and electrostatic potential distribution for the single flexible polyelectrolyte chain in ``free ends'' $[(a)$ and $(b)$, respectively $]$ and ``end-fixed'' state $[(c)$ and $(d)$, respectively$]$ at a higher monomer volume fraction compared to Fig. 2. Parameters used to generate these plots are $l_{B}/b = 3, \alpha = 0.1, c_{s} = 0.1M, N = 100, R/b = 4$ and $\chi_{ps} = 0.45$. For plots $(c)$ and $(d)$, one end is anchored at $\left[x,y\right] = \left[(R-0.625)b,0\right]$ (shown by arrow).
\end{description}

\begin{description}
\item[Fig. 5.:] (Color) Effect of confinement on the monomer density distribution along $x$ 
and $y$ axes for the single flexible polyelectrolyte chain in ``free ends'' $[(a)$ and $(b)$, respectively $]$ and ``end-fixed'' state $[(c)$ and $(d)$, respectively$]$. Parameters used to generate these plots are $l_{B}/b = 3, \alpha = 0.1, c_{s} = 0.1M, R/b = 4$ and $\chi_{ps} = 0.45$. For plots $(c)$ and $(d)$, one end is anchored at $\left[x,y\right] = \left[(R-0.625)b,0\right]$.
\end{description}

\begin{description}
\item[Fig. 6.:] Comparison of the monomer density distribution for the single flexible polyelectrolyte chain in ``free ends'' and ``end-fixed'' states. Figs. $(a)$ and $(b)$ correspond to the density distribution along $x$ and $y$ axes, respectively for $N=50$. 
Similarly, Figs. $(c)$ and $(d)$ correspond to the density distribution along $x$ and $y$ axes, respectively for $N=100$. All the other parameters are the same as in Fig. ~\ref{fig:den_pot_n50} i.e., $l_{B}/b = 3, \alpha = 0.1, c_{s} = 0.1M, R/b = 4$ and $\chi_{ps} = 0.45$. For the ``end-fixed'' state in plots, one end is anchored at $\left[x,y\right] = \left[(R-0.625)b,0\right]$.
\end{description}

\begin{figure}[ht!]
\vspace*{-0.95cm}
\begin{center}
(a)$\begin{array}{c@{\hspace{.001in}}c@{\hspace{.001in}}}
\includegraphics[width=3in,height=3in]{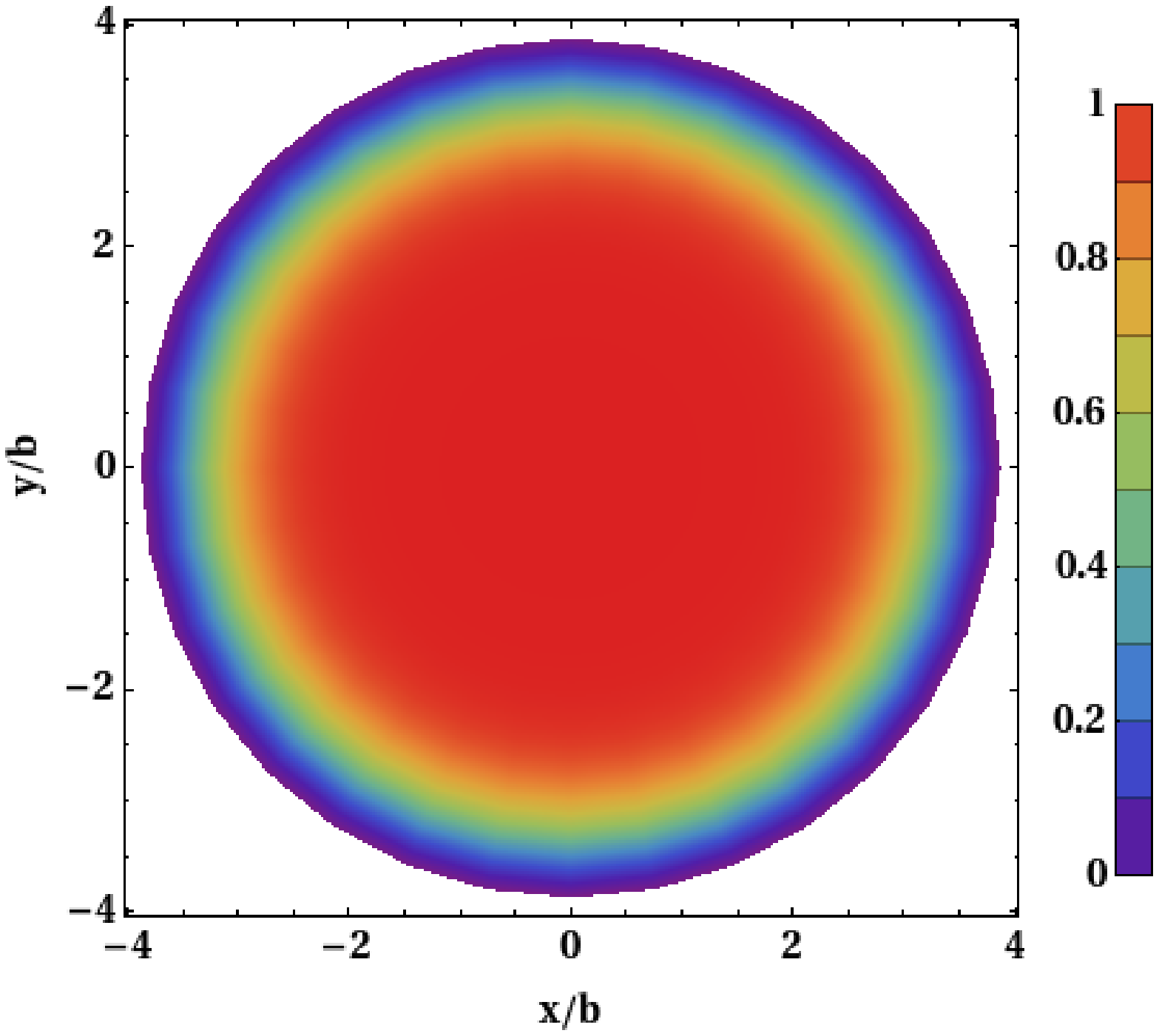}  &
\includegraphics[width=3in,height=3in]{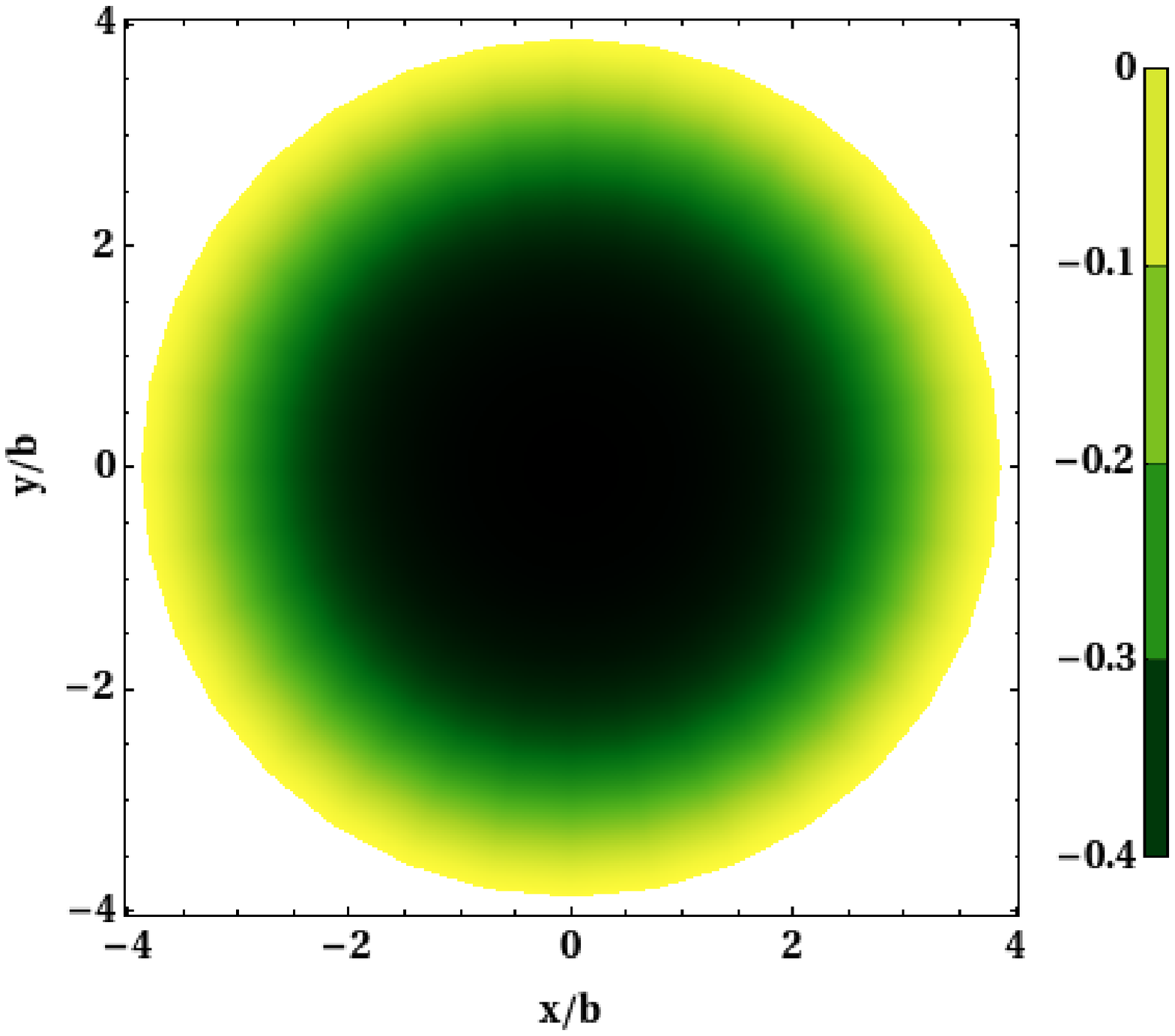}
\end{array}$(b)
\newline
(c)$\begin{array}{c@{\hspace{.001in}}c@{\hspace{.001in}}}
 \includegraphics[width=3in,height=3in]{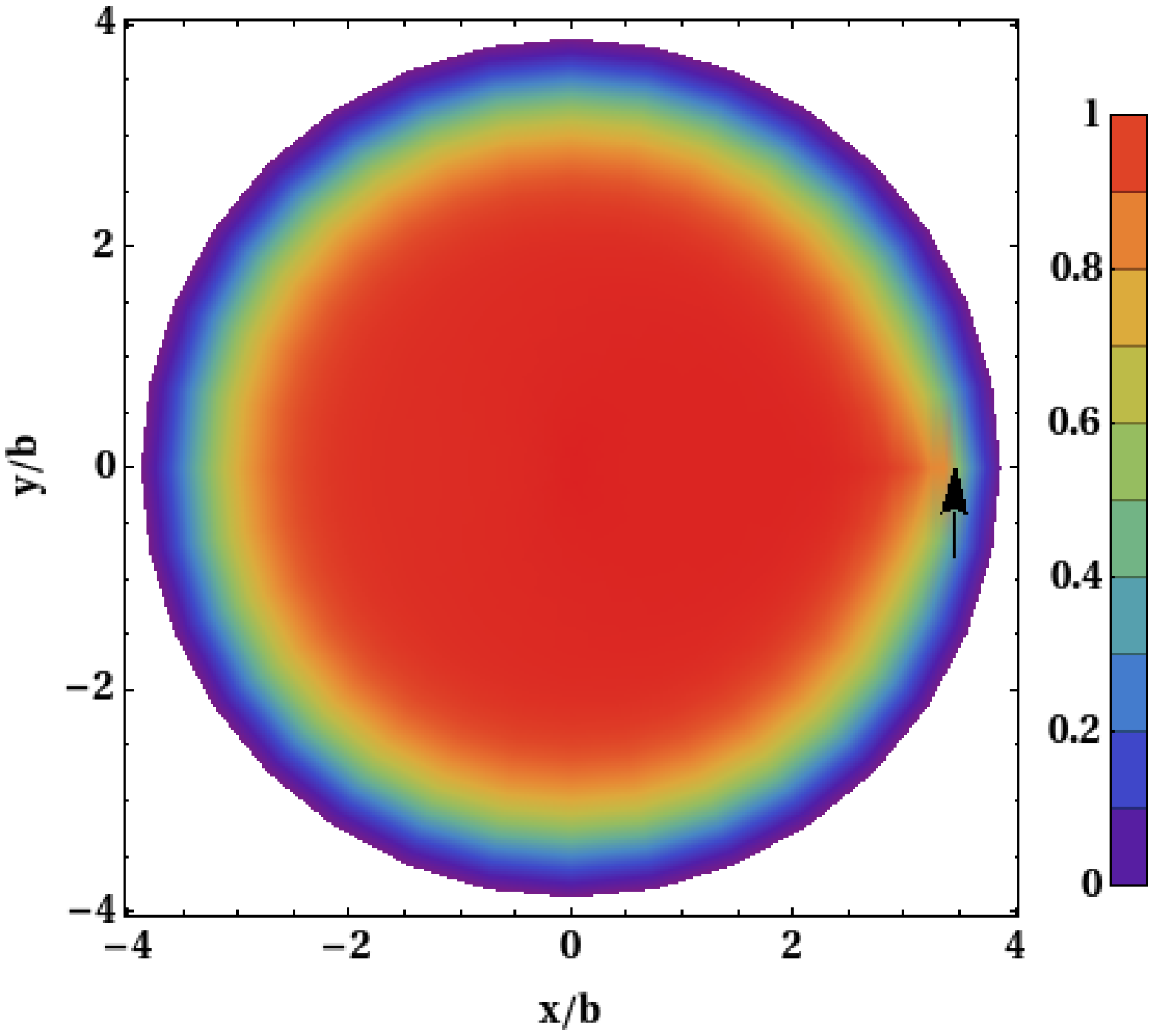} &
 \includegraphics[width=3in,height=3in]{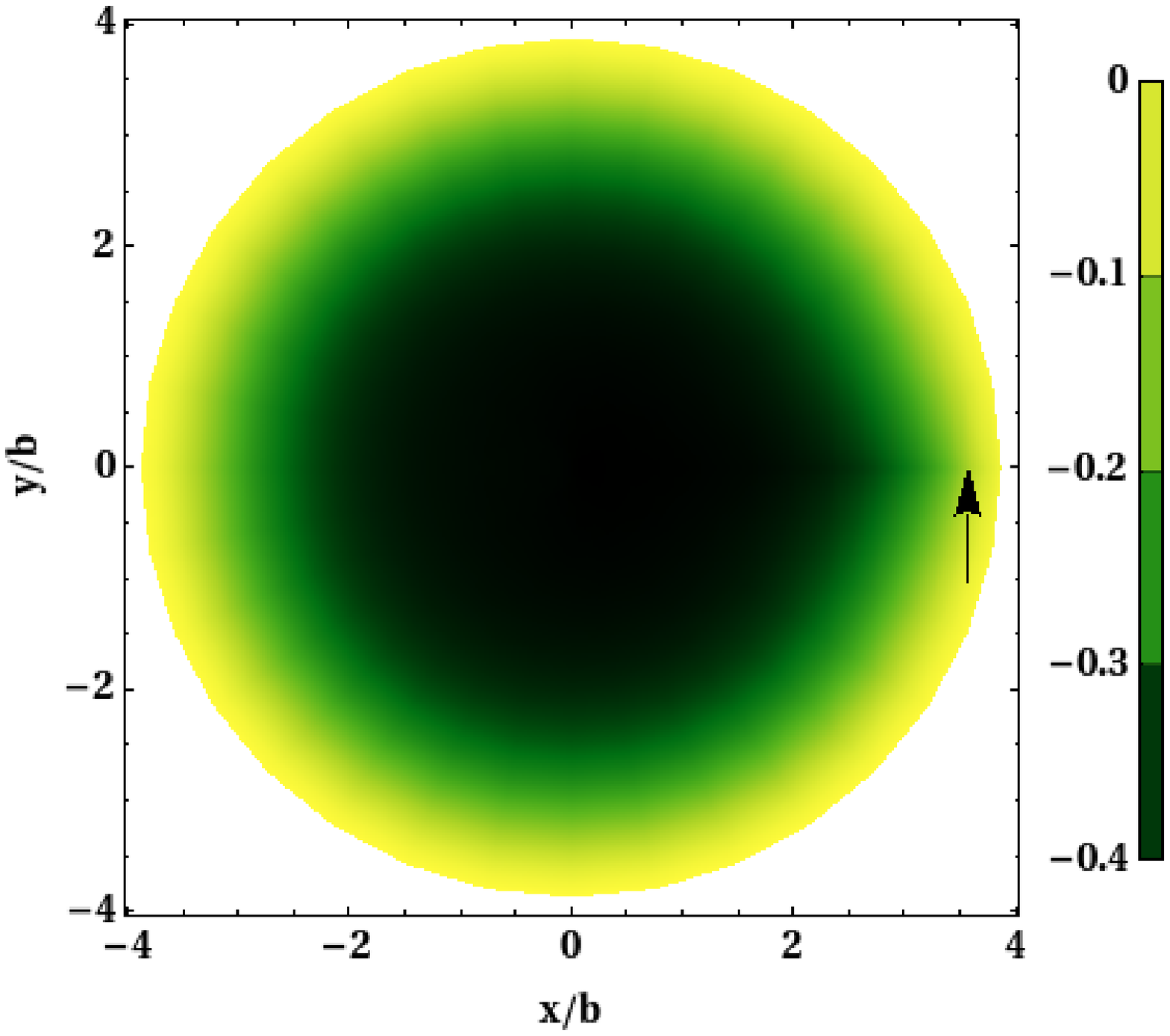}
\end{array}$(d)
\vspace*{-1.0cm}
\newline
\end{center}
\caption{(Color) Monomer and electrostatic potential distribution for the single flexible polyelectrolyte chain in ``free ends'' $[(a)$ and $(b)$, respectively $]$ and ``end-fixed'' state $[(c)$ and $(d)$, respectively$]$ at a higher monomer volume fraction compared to Fig. 2. Parameters used to generate these plots are $l_{B}/b = 3, \alpha = 0.1, c_{s} = 0.1M, N = 100, R/b = 4$ and $\chi_{ps} = 0.45$. For plots $(c)$ and $(d)$, one end is anchored at $\left[x,y\right] = \left[(R-0.625)b,0\right]$ (shown by arrow).} \label{fig:den_pot_n100}
\end{figure}

\begin{figure}[ht!]
\vspace*{-0.95cm}
\begin{center}
(a)$\begin{array}{c@{\hspace{.001in}}c@{\hspace{.001in}}}
\includegraphics[width=3in,height=3in]{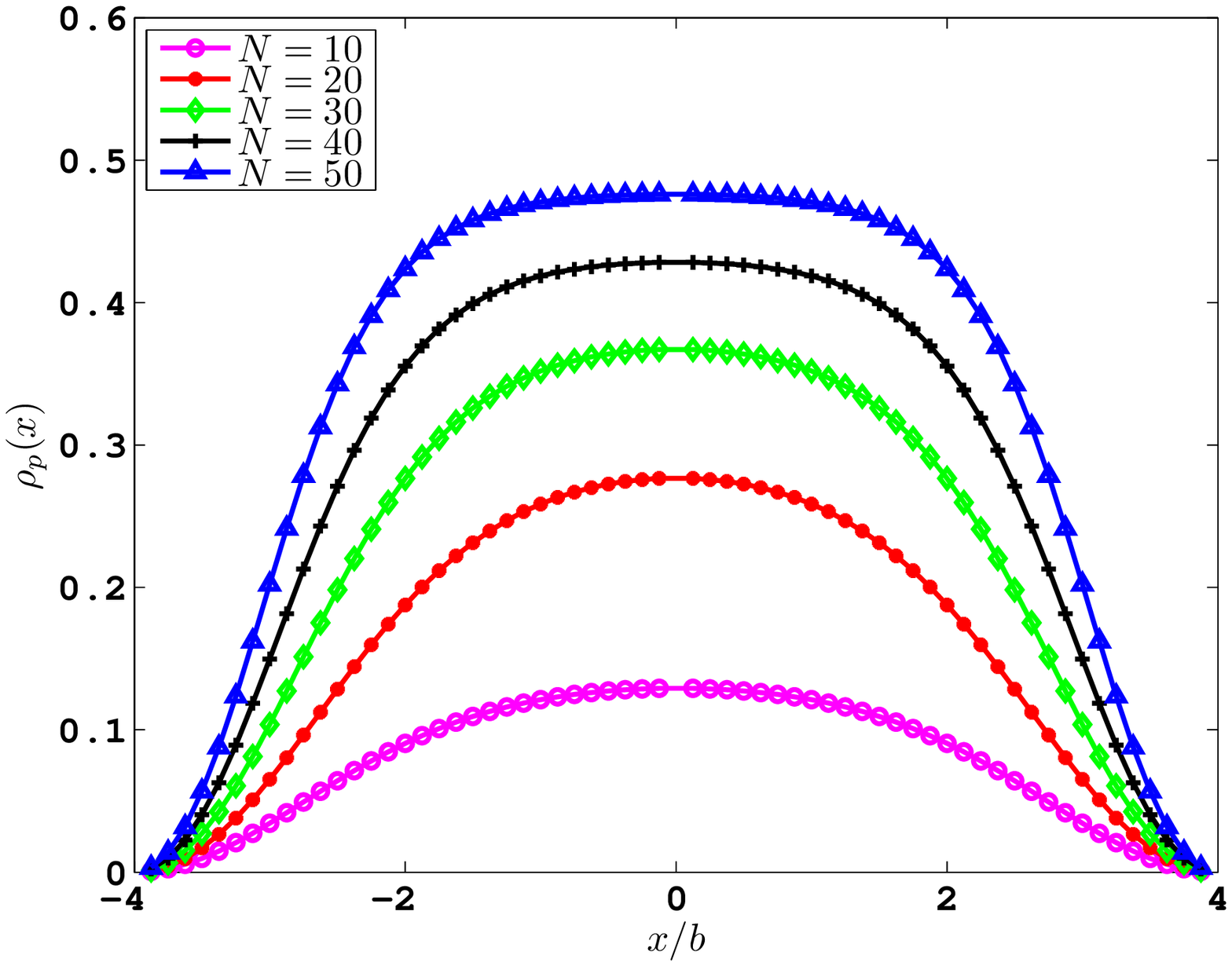}  &
\includegraphics[width=3in,height=3in]{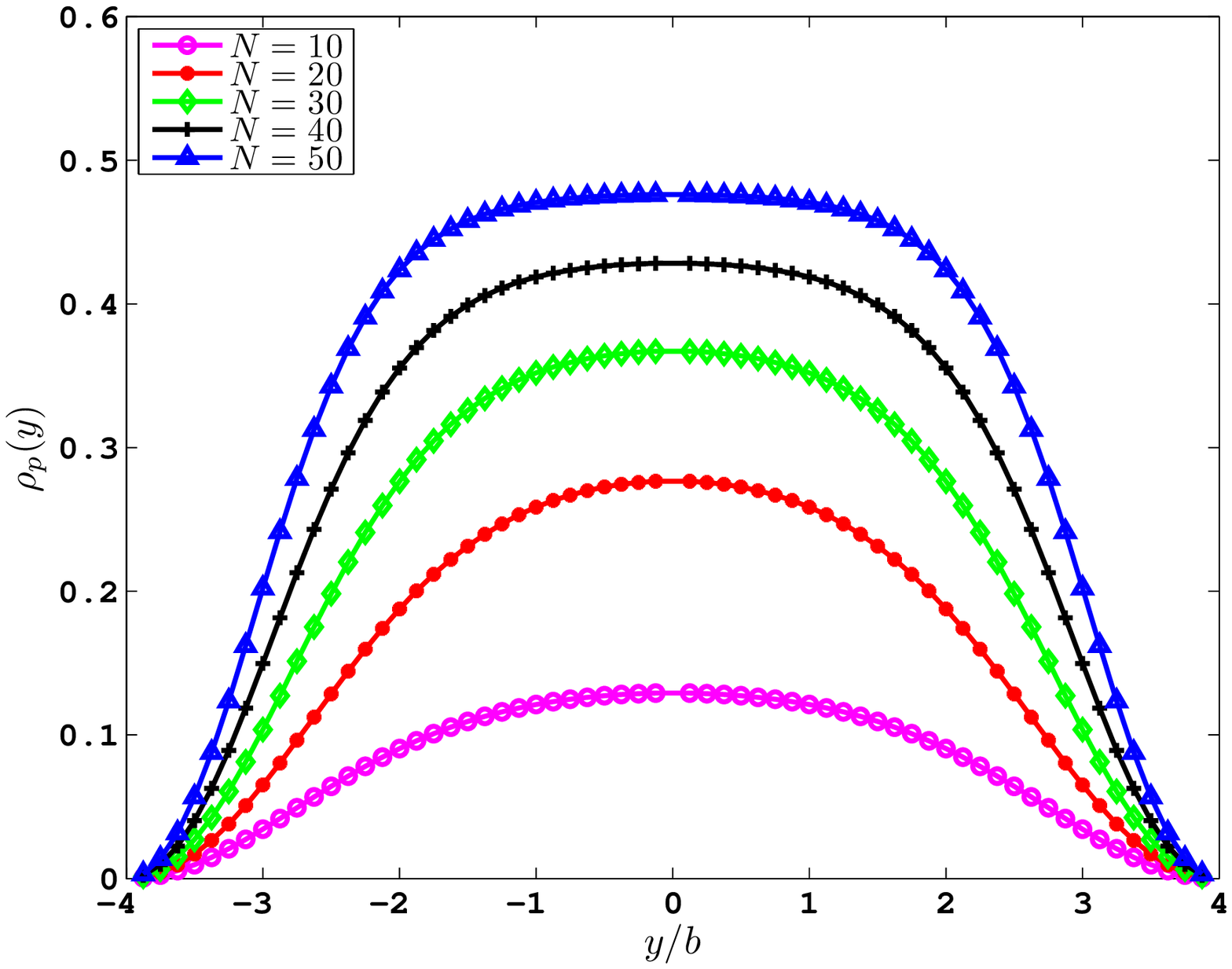}
\end{array}$(b)
\newline
(c)$\begin{array}{c@{\hspace{.001in}}c@{\hspace{.001in}}}
 \includegraphics[width=3in,height=3in]{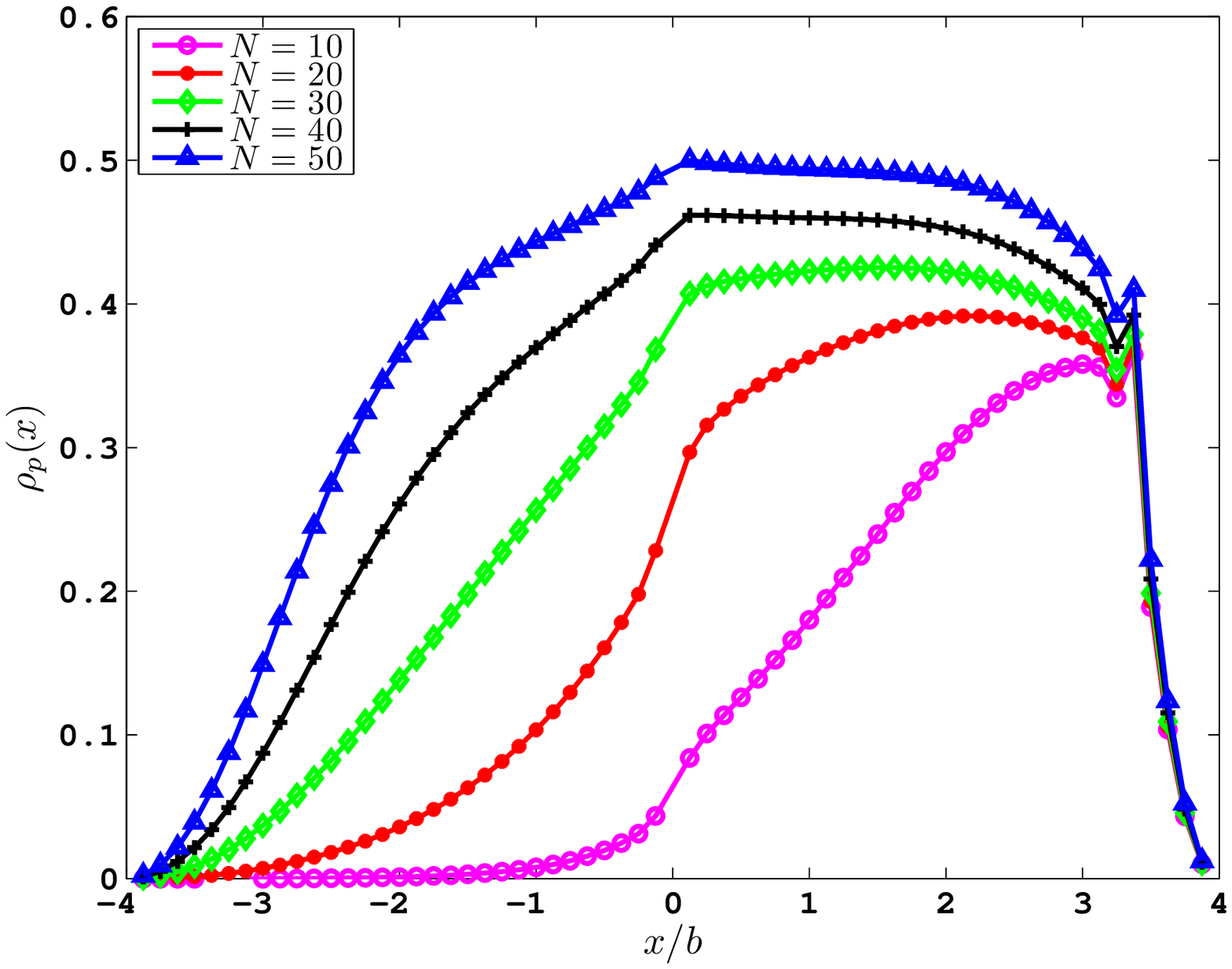} &
 \includegraphics[width=3in,height=3in]{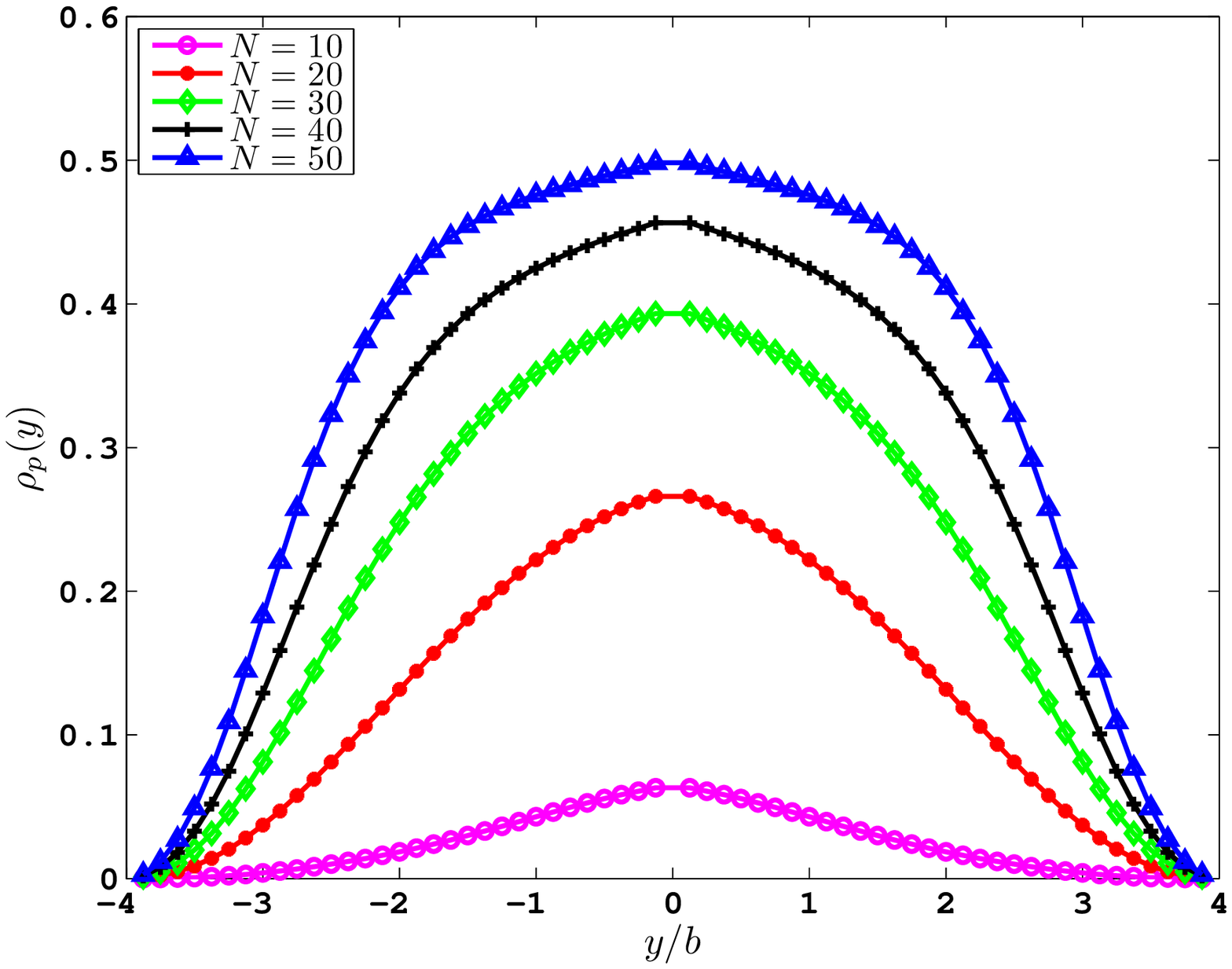}
\end{array}$(d)
\vspace*{-1.0cm}
\newline
\end{center}
\caption{(Color) Effect of confinement on the monomer density distribution along $x$ 
and $y$ axes for the single flexible polyelectrolyte chain in ``free ends'' $[(a)$ and $(b)$, respectively $]$ and ``end-fixed'' state $[(c)$ and $(d)$, respectively$]$. Parameters used to generate these plots are $l_{B}/b = 3, \alpha = 0.1, c_{s} = 0.1M, R/b = 4$ and $\chi_{ps} = 0.45$. For plots $(c)$ and $(d)$, one end is anchored at $\left[x,y\right] = \left[(R-0.625)b,0\right]$.} \label{fig:den_all_n}
\end{figure}

\begin{figure}[ht!]
\vspace*{-0.95cm}
\begin{center}
$\begin{array}{c@{\hspace{.001in}}c@{\hspace{.001in}}}
\includegraphics[width=2.75in,height=2.75in]{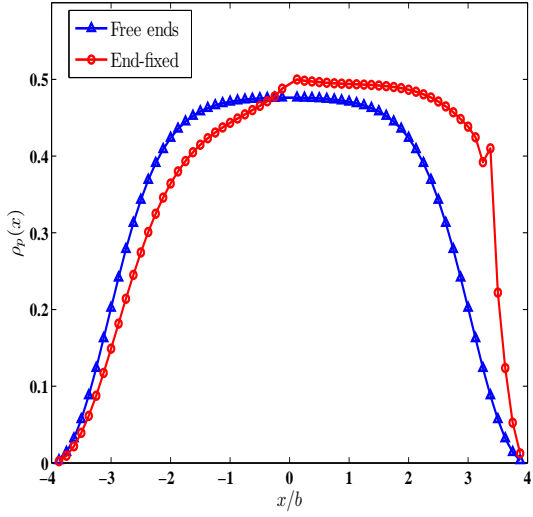}\hspace*{0.5cm}  &
\includegraphics[width=2.75in,height=2.75in]{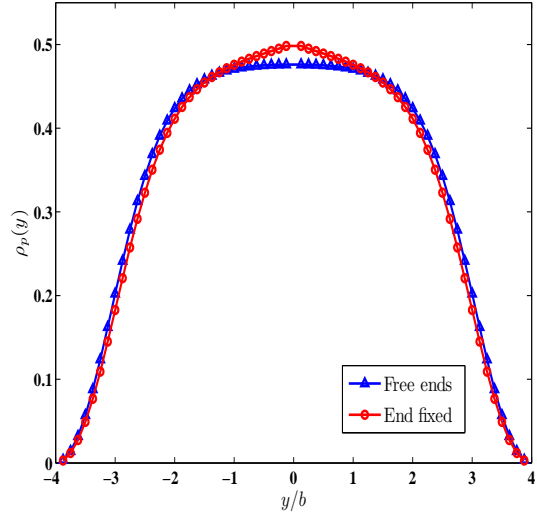}\hspace*{0.5cm}\\
\textbf{(a)} & \textbf{(b)} \\
   & \\
   \includegraphics[width=2.75in,height=2.75in]{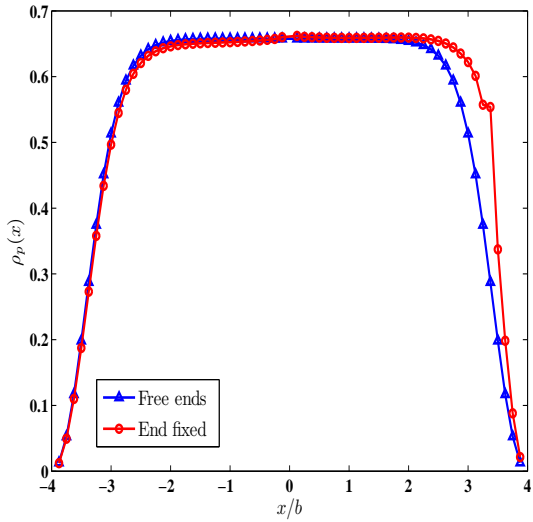}\hspace*{0.5cm} &
\includegraphics[width=2.75in,height=2.75in]{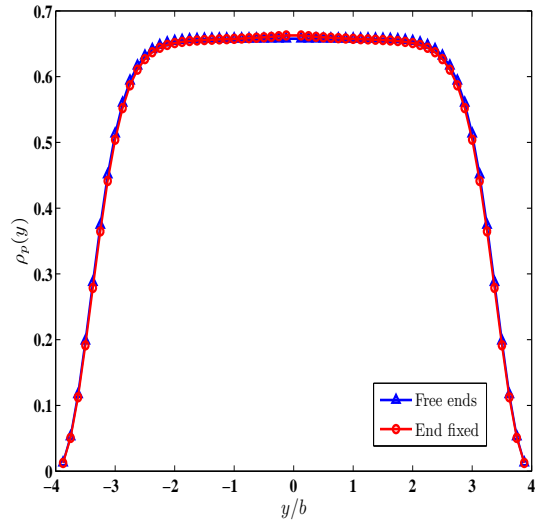}\hspace*{0.5cm}\\
\textbf{(c)} & \textbf{(d)} 
\end{array}$
\vspace*{-1.0cm}
\newline
\end{center}
\caption{Comparison of the monomer density distribution for the single flexible polyelectrolyte chain in ``free ends'' and ``end-fixed'' states. Figs. $(a)$ and $(b)$ correspond to the density distribution along $x$ and $y$ axes, respectively for $N=50$. 
Similarly, Figs. $(c)$ and $(d)$ correspond to the density distribution along $x$ and $y$ axes, respectively for $N=100$. All the other parameters are the same as in Fig. ~\ref{fig:den_pot_n50} i.e., $l_{B}/b = 3, \alpha = 0.1, c_{s} = 0.1M, R/b = 4$ and $\chi_{ps} = 0.45$. For the ``end-fixed'' state in plots, one end is anchored at $\left[x,y\right] = \left[(R-0.625)b,0\right]$.} \label{fig:densities_x_y_n}
\end{figure}

\begin{figure}[ht!]
\vspace*{-0.95cm}
\begin{center}
$\begin{array}{c@{\hspace{.001in}}c@{\hspace{.001in}}}
\includegraphics[width=3.5in,height=3.5in]{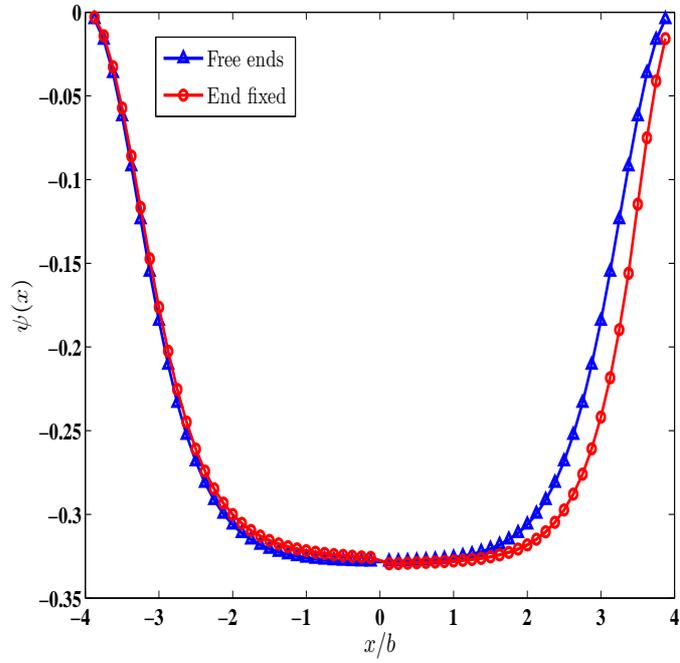}&   \\
\textbf{(a)} & \\
& \\
\includegraphics[width=3.5in,height=3.5in]{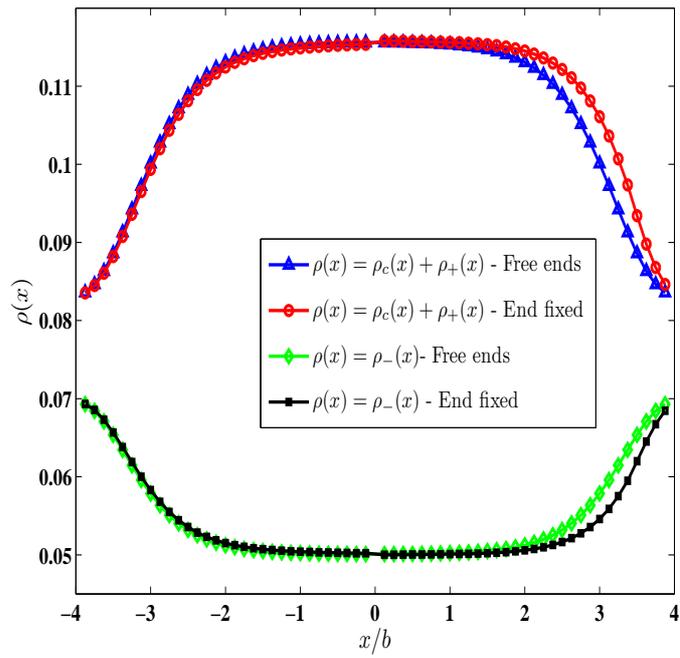} & \\
\textbf{(b)} &
\end{array}$
\vspace*{-1.0cm}
\newline
\end{center}
\caption{Comparison of the electrostatic potential $(a)$, and the counterion ($\rho_c (x) + \rho_+(x)$) and coion ($\rho_{-}(x)$) density distributions $(b)$ for the single flexible polyelectrolyte chain in ``free ends'' and ``end-fixed'' state, respectively. All the parameters are the same as in Fig. ~\ref{fig:den_pot_n100} 
i.e., $l_{B}/b = 3, \alpha = 0.1, c_{s} = 0.1M, R/b = 4, N = 100$ and $\chi_{ps} = 0.45$. } \label{fig:ions_x_n}
\end{figure}

\begin{description}
\item[Fig. 7.:] Comparison of the electrostatic potential $(a)$, and the counterion ($\rho_c (x) + \rho_+(x)$) and coion ($\rho_{-}(x)$) density distributions $(b)$ for the single flexible polyelectrolyte chain in ``free ends'' and ``end-fixed'' state, respectively. All the parameters are the same as in Fig. ~\ref{fig:den_pot_n100} 
i.e., $l_{B}/b = 3, \alpha = 0.1, c_{s} = 0.1M, R/b = 4, N = 100$ and $\chi_{ps} = 0.45$. 
\end{description}

\begin{description}
\item[Fig. 8.:] (Color) Effect of $N$ and $R$ on free energy barriers for the chain end to be localized on the surface of a neutral spherical cavity. $l_{B}/b = 3, \alpha = 0.1, c_{s} = 0.1M$ and $\chi_{ps} = 0.45$.
\end{description}

\begin{figure}[ht!]
\includegraphics[width=5in,height=5in]{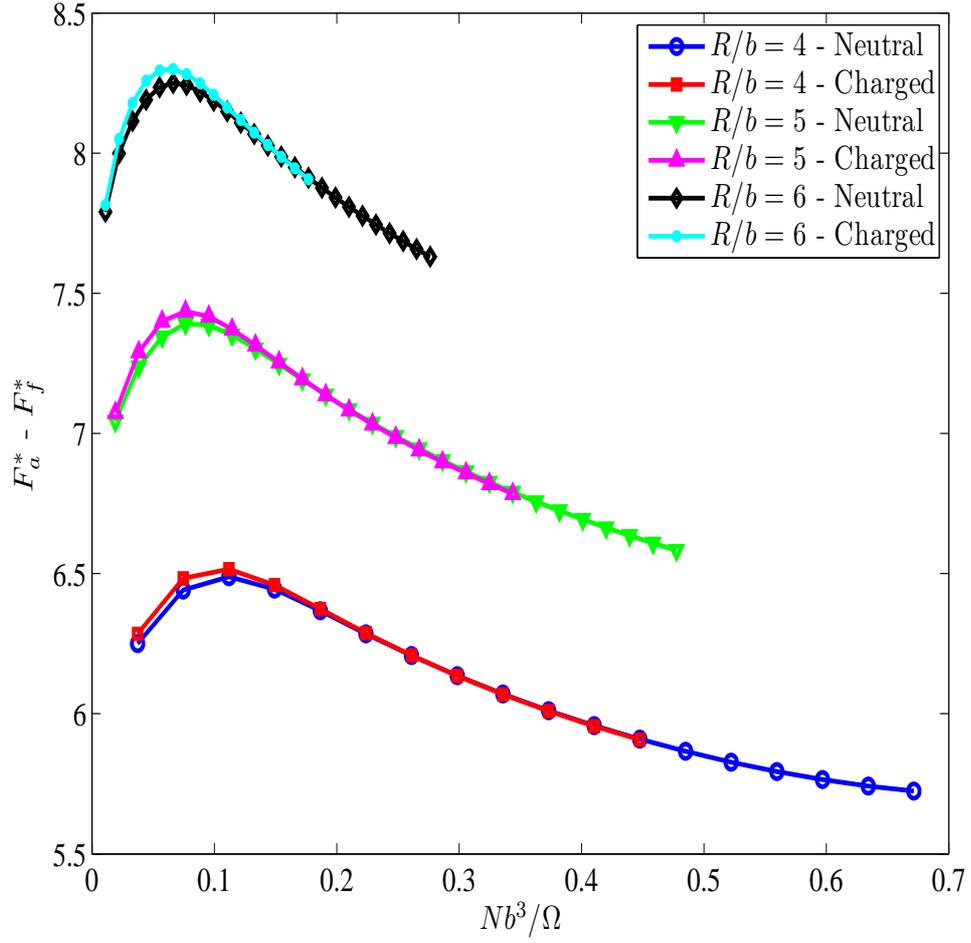}
\caption{(Color) Effect of $N$ and $R$ on free energy barriers for the chain end to be localized on the surface of a neutral spherical cavity. $l_{B}/b = 3, \alpha = 0.1, c_{s} = 0.1M$ and $\chi_{ps} = 0.45$.}\label{fig:barriers_r}
\end{figure}

\begin{figure}[ht!]
\vspace*{-0.95cm}
\begin{center}
$\begin{array}{c@{\hspace{.001in}}c@{\hspace{.001in}}}
\includegraphics[width=3.5in,height=3.5in]{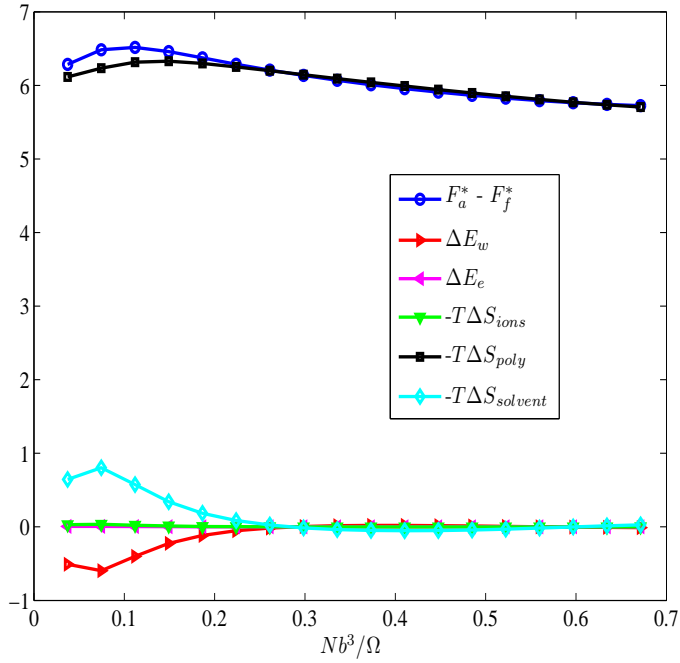}&   \\
\textbf{(a)} & \\
& \\
\includegraphics[width=3.5in,height=3.5in]{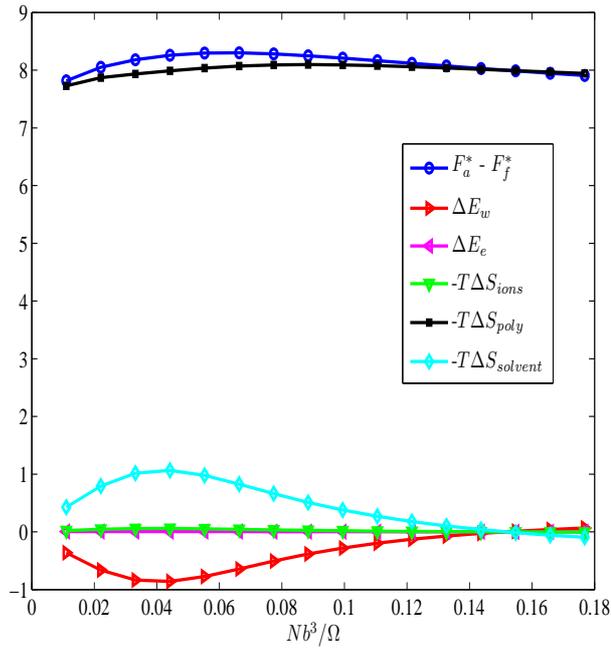} & \\
\textbf{(b)} &
\end{array}$
\vspace*{-1.0cm}
\newline
\end{center}
\caption{(Color) Dominance of conformational entropy to free energy barriers. For these figures, $l_{B}/b = 3, \alpha = 0.1, c_{s} = 0.1M, \chi_{ps} = 0.45,$ and Figs. (a) and (b) correspond to $R/b = 4$ and $R/b = 6$, respectively. The net free energy barriers in the figures are the same as in Fig.  ~\ref{fig:barriers_r}.} \label{fig:barrier_comps}
\end{figure}

\begin{figure}[ht!]
\includegraphics[width=5in,height=5in]{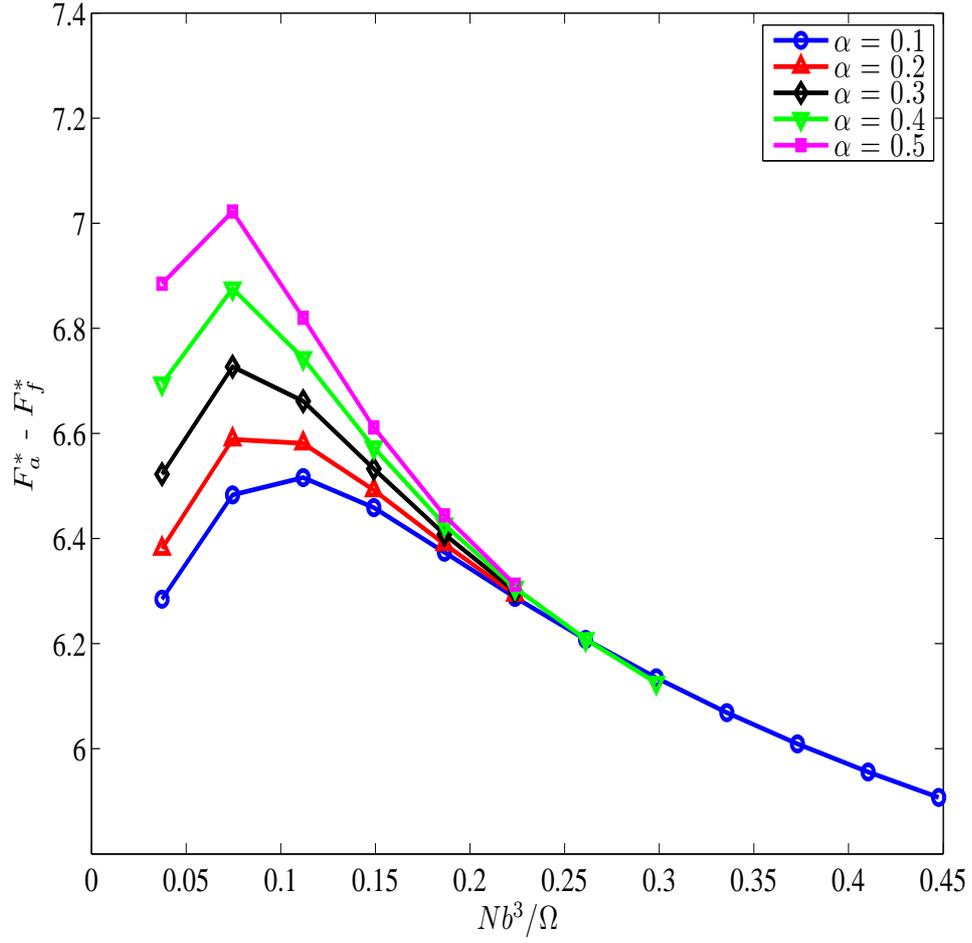}
\caption{(Color) Dependence of the free energy barriers on the degree of ionization of  the polyelectrolyte chain. Parameters used to obtain these plots are: $l_{B}/b = 3, c_{s} = 0.1M, R/b = 4,\chi_{ps} = 0.45$. } \label{fig:alpha_effect}
\end{figure}

\begin{figure}[ht!]
\includegraphics[width=5in,height=5in]{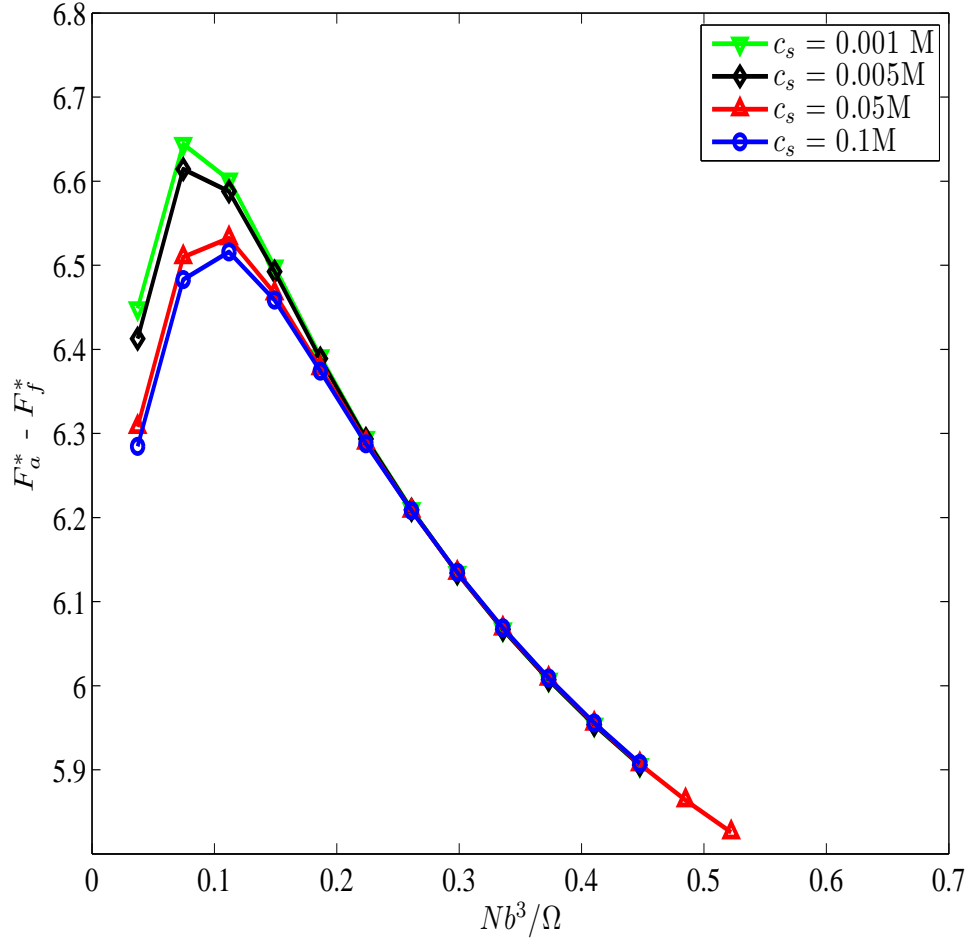}
\caption{(Color) Effect of the added salt concentration on the free energy barriers. Parameters used to obtain these plots are: $l_{B}/b = 3, \alpha = 0.1, R/b = 4,\chi_{ps} = 0.45$. } \label{fig:cs_effect}
\end{figure}

\begin{figure}[ht!]
\includegraphics[width=5in,height=5in]{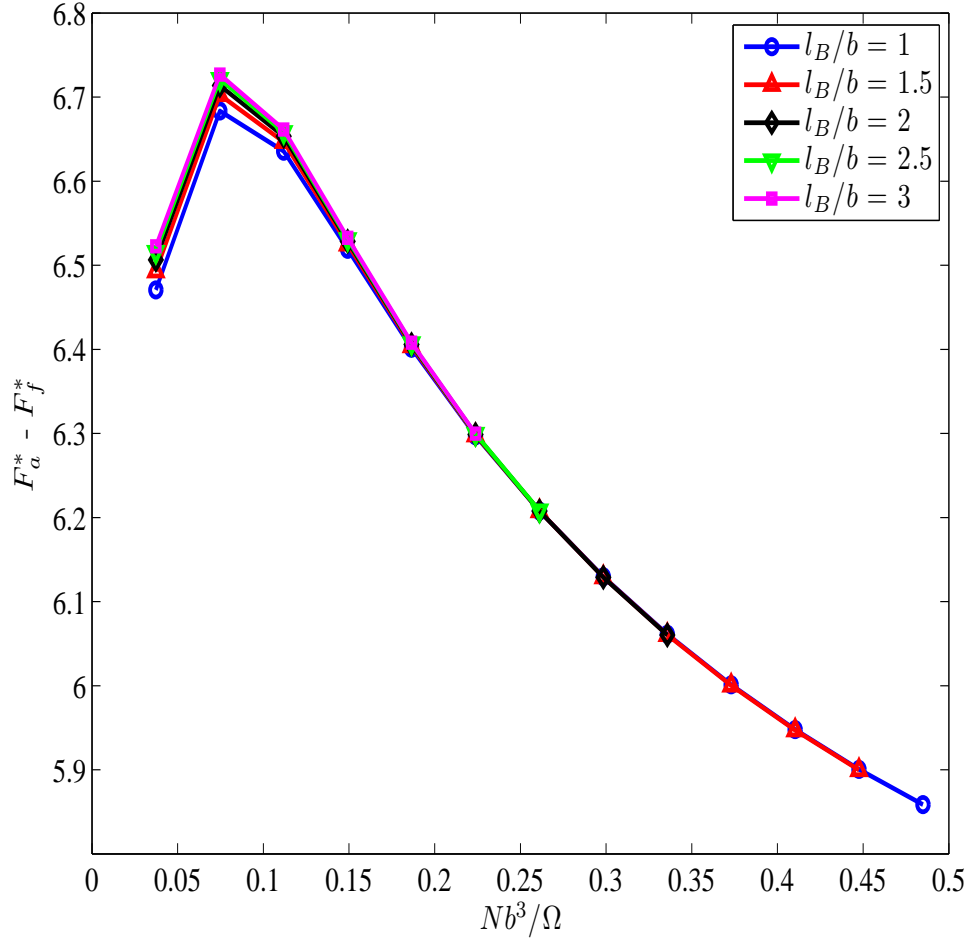}
\caption{(Color) Dependence of the free energy barriers on Bjerrum length, which characterizes the electrostatic interaction strength between charged species. Parameters used to obtain these plots are: $\alpha = 0.3, c_{s} = 0.1M, R/b = 4,\chi_{ps} = 0.45$. } \label{fig:lb_effect}
\end{figure}

\begin{figure}[ht!]
\includegraphics[width=5in,height=5in]{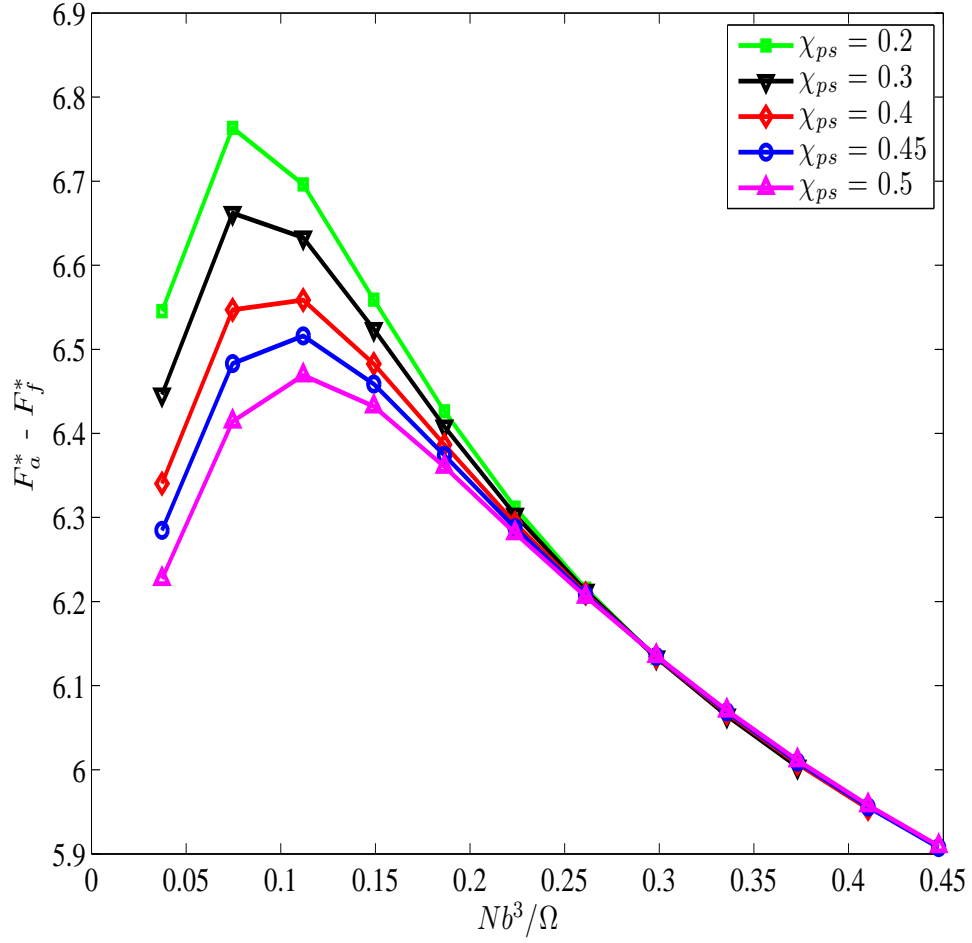}
\caption{(Color) Effect of Flory's chi parameter (characterizing the exlcuded 
volume interactions between monomers and solvent molecules) on the free energy barriers is shown here. Parameters used to obtain these plots are: $l_{B}/b = 3, \alpha = 0.1, c_{s} = 0.1M, R/b = 4$. } \label{fig:chi_effect}
\end{figure}

\begin{description}
\item[Fig. 9.:] (Color) Dominance of conformational entropy to free energy barriers. For these figures, $l_{B}/b = 3, \alpha = 0.1, c_{s} = 0.1M, \chi_{ps} = 0.45,$ and Figs. (a) and (b) correspond to $R/b = 4$ and $R/b = 6$, respectively. The net free energy barriers in the figures are the same as in Fig.  ~\ref{fig:barriers_r}.
\end{description}

\begin{description}
\item[Fig. 10.:] (Color) Dependence of the free energy barriers on the degree of ionization of  the polyelectrolyte chain. Parameters used to obtain these plots are: $l_{B}/b = 3, c_{s} = 0.1M, R/b = 4,\chi_{ps} = 0.45$. 
\end{description}

\begin{description}
\item[Fig. 11.:] (Color) Effect of the added salt concentration on the free energy barriers. Parameters used to obtain these plots are: $l_{B}/b = 3, \alpha = 0.1, R/b = 4,\chi_{ps} = 0.45$. 
\end{description}

\begin{description}
\item[Fig. 12.:] (Color) Dependence of the free energy barriers on Bjerrum length, which characterizes the electrostatic interaction strength between charged species. Parameters used to obtain these plots are: $\alpha = 0.3, c_{s} = 0.1M, R/b = 4,\chi_{ps} = 0.45$. 
\end{description}

\begin{description}
\item[Fig. 13.:] (Color) Effect of Flory's chi parameter (characterizing the exlcuded 
volume interactions between monomers and solvent molecules) on the free energy barriers is shown here. Parameters used to obtain these plots are: $l_{B}/b = 3, \alpha = 0.1, c_{s} = 0.1M, R/b = 4$. 
\end{description}

\newpage
\newpage
\newpage

\newpage

\newpage

\newpage

\newpage

\newpage

\newpage

\newpage

\newpage

\newpage

\newpage


\end{document}